# Explicit three dimensional topology optimization via Moving Morphable Void (MMV) approach

Weisheng Zhang[1], Jishun Chen[1], Xuefeng Zhu[2], Jianhua Zhou[1], Dingchuan Xue[1], Xin Lei[1] and Xu Guo[1]*

[1]*State Key Laboratory of Structural Analysis for Industrial Equipment,*

*Department of Engineering Mechanics,*

*International Research Center for Computational Mechanics,*

*Dalian University of Technology, Dalian, 116023, P.R. China*

[2]*School of Automotive Engineering,*

*Dalian University of Technology, Dalian, 116023, P.R. China*

## Abstract

Three dimensional (3D) topology optimization problems always involve huge numbers of Degrees of Freedom (DOFs) in finite element analysis (FEA) and design variables in numerical optimization, respectively. This will inevitably lead to large computational efforts in the solution process. In the present paper, an efficient and explicit topology optimization approach which can reduce not only the number of design variables but also the number of degrees of freedom in FEA is proposed based on the Moving Morphable Voids (MMVs) solution framework. This is achieved by introducing a set of geometry parameters (e.g., control points of B-spline surfaces) to describe the boundary of a structure explicitly and removing the unnecessary DOFs from the FE model at every step of numerical optimization. Numerical examples demonstrate that the proposed approach does can overcome the bottleneck problems associated with a 3D topology optimization problem in a straightforward way and enhance the solution efficiency significantly.

*Corresponding author. E-mail: guoxu@dlut.edu.cn Tel: +86-411-84707807

# 1. Introduction

Structural topology optimization aims at distributing a given amount of material in a prescribed design domain in an optimal way in order to get optimized structural performances. Topology optimization has undergone rapid development since the pioneering works of Prager and Rozvany [1], Cheng and Olhoff [2], Bendsøe and Kikuchi [3] and Zhou and Rozvany [4]. Nowadays, numerous powerful methods have been proposed for structural topology optimization and some of them have already found successful applications in many application fields. We refer the readers to [5-8] and the references therein for a state-of-the-art reviews of the recent developments in topology optimization.

Compared to two dimensional topology optimization (2D TopOpt), three dimensional topology optimization (3D TopOpt) is more challenging from computational point of view especially when traditional solution approaches (e.g., variable density approach [9, 10] and level set approach [11, 12]) are adopted. This can be explained as follows. For example, if a $1 \times 1 \times 1$ cube design domain (shown in Fig. 1) is discretized by 100 elements along each coordinate direction (note that this is actually not a high resolution), then both the number of DOFs (i.e, $nd$) in finite element analysis (FEA) and the number of design variables (i.e, $nv$) associated with numerical optimization will reach the order of one million when the aforementioned approaches are employed! This will inevitably lead to large computational demand since the computational complexities for both FEA and numerical optimization will increase almost cubically with respect to the number of $nd$ and $nv$, respectively [13]. The computational challenge will become more serious when optimal solutions with higher resolutions are required.

Recent years witnessed a growing interest on optimizing the topology of three dimensional problems [14] which are ubiquitous in practical engineering applications. For example, Diaz and Lipton [15] investigated the 3D TopOpt problems using a moment formulation to characterize the optimal orthotropy of layered material at each point of the design domain. The optimal material distribution in 3D space was then found through a two-level approach where the global material density field and the local moment variables were optimized in a hierarchical way. Fernandes et al. [16] proposed a modified homogenization approach to optimize the topology of 3D linear elastic structures. A constraint on the ‘perimeter’ of the structure is introduced to guarantee the well-posedness of the problem formulation. In [17], Villanueva and Maute developed a novel approach for 3D TopOpt problems by combining the level set method (LSM) and eXtended Finite Element Method (XFEM). Numerical results indicated that optimized 3D structures with crisp boundaries can be obtained with use of relatively coarse FE meshes. Aage and Lazarov [18] discussed how to solve large scale 3D TopOpt problems using modern parallel computation techniques. Furthermore, an efficient Matlab code for 3D TopOpt was also provided in [19]. It is worth noting, however, that the bottleneck problem associated with 3D TopOpt, that is reducing the number of DOFs in FEA and the number of design variables associated with numerical optimization had not been well resolved in the above works. Recently, in [20], a new approach for 3D TopOpt is proposed based on the so-called Moving Morphable Component (MMC) approach where a set of moving morphable components are adopted as basic building blocks for topology optimization and optimal structural topology can be obtained by optimizing the shapes, orientations and layout of the components [21] . The attractive feature of this approach is that the number of design variables involved in the problem formulation can be reduced substantially. This is very helpful for reducing the computational time associated with the numerical optimization process. Another challenging issue for 3D TopOpt that is the large number of DOFs in FEA, however, had not been addressed in this work.

The objective of the present paper is to propose a 3D TopOpt approach under the newly developed Moving Morphable Voids (MMVs) solution framework where a set of moving morphable voids are adopted as basic building blocks for topology optimization [22, 23]. In [23] and [23], the authors only considered 2D problems and only the potential of the MMV approach for reducing the computational effort from *numerical optimization aspect* had been demonstrated. As will be clear in the following discussions, the MMV approach can not only reduce the number of design variables involve in a TopOpt problem, but also the number of DOFs in FEA, which constitutes the main bottleneck for the solution efficiency of a 3D problem. In order to fully explore the potential of the MMV approach for enhancing the solution efficiency of TopOpt problem, in the present paper, the MMV approach will be extended to solve the challenging bottleneck problems associated with 3D TopOpt problems by introducing some novel ideas and techniques.

The remainder of the paper is organized as follows. In Section 2, first the basic idea of the MMV approach where topology optimization can be achieved by varying the boundary of the structure explicitly is introduced briefly. Section 3 is devoted to explaining how to describe the shape of a 3D structure through a set of geometry parameters. The mathematical formulation for 3D TopOpt problems under the MMV-based solution framework will also be presented in this section. Section 4 discusses the numerical implementation issue in details. In Section 5, the effectiveness of the proposed approach is demonstrated through three examples. Comparisons of solution efficiencies with those of traditional methods are also made in this section. Finally, some concluding remarks are provided in Section 6.

## 2. Moving Morphable Void (MMV) approach

As pointed out in [23], MMV approach is actually a dual version of the Moving Morphable Component (MMC) approach first proposed by Guo et al. in [21] where topology optimization can be achieved in an explicit and geometrical way. In contrast to the MMC approach, where a set of moving morphable components are adopted as the building blocks of topology optimization, a set of moving morphable voids are used to optimize the structural topology in the MMV approach. In the MMV approach, as shown in Fig. 2, the topological change of a structure is achieved by the deformation, intersection and merging of a set of closed parametric curves (for 2D problems) or parametric surfaces (for 3D problems) that represent the boundary of the structure. Unlike the traditional level set approach, there is no need to introduce an extra level set function defined in a higher dimensional space to represent the structural boundary implicitly in the MMV approach. The design variables involved in the MMV approach are only the coordinates of the control points (or some related parameters) of the parametric curves/surfaces. This greatly reduced the number of design variables involved in the optimization problem. The readers are referred to [23] for more discussions on technical details of this approach.

As discussed intensively in [23], one of the key points in MMV approach is to describe the geometries of the morphable voids in an explicit and parametric way. The adopted geometry description scheme should also be computationally amenable. In the following section, we will discuss how to describe the geometry of a 3D structure through a set of parameters in the MMV-based solution framework.

## 3. Geometry description of a 3D structure in the MMV approach

Let us denote the 3D design domain where optimal material distribution is sought for as $D$. Then the region occupied by the optimized structure can be described as $\Omega^s = \mathrm{D} \backslash \bigcup_{i=1}^{n} \Omega^i$, where $\Omega^i,\ i = 1, \ldots, n$ are a set of void regions enclosed by a set of closed surfaces $S^i,\ i = 1, \ldots, n$ (see Fig. 3 for a schematic illustration). With use of this geometry description, the optimal structural topology of a 3D structure can be uniquely determined by optimizing the positions, shapes and layout of $S^i,\ i = 1, \ldots, n$.

***Remark 1:*** As pointed out in [23], in topology optimization it is highly desirable to have a global, explicit and smooth description of the boundary of an optimized structure for the purpose of establishing a direct link between the optimization result and CAD/CAE systems. For 2D topology optimization problems, this had already been realized by the approach presented in [23] where a smooth B-spline description can be constructed by rearranging the control points of a set of smooth skeleton curves for every single piece of structural boundary (we refer the readers to [23] for more technical details). For 3D problems, constructing a global, explicit and smooth description of a boundary surface from individual skeleton surfaces is, however, a more challenging task since under this circumstance it is difficult to establish a sequential order for control points on the involved skeleton surfaces as in 2D case. In the present study, we shall neglect the smoothness and global explicit description issues of structural boundary and intend to resolve these challenging problems in a separate work.

Next, let us consider how to describe the shape of a smooth surface in a way that is suitable for carrying out topology optimization efficiently. Actually, if adaptive mesh is employed for FEA, any approach that can give an explicit description of boundary surface can be used to construct the FE mesh directly. If, however, fixed mesh is preferable for FEA as in traditional topology optimization methods, the so-called topology description function (TDF) is an appropriate tool to establish the link between the structural topology and FE model of the structure [24]. Actually, the construction of parameterized TDFs have been studied intensively in the field of computer graphics [25-27]. One of the challenging issues is to develop the TDF construction techniques which are applicable not only to star-shaped regions but also to non-star-shaped regions. Here the so-called star-shaped region refers to a domain where a central interior point from which the entire boundary is visible can be found. We also refer the readers to Fig. 4 for a schematic illustration of the star-shaped and non-star-shaped regions, respectively. Under the MMC-based solution framework, a unified approach, which is applicable for constructing the TDFs of both star-shaped and non-star-shaped regions systematically, is proposed. In this approach, as shown in Fig. 5, the skeleton (medial line for 2D and medial surface for 3D) of a component and the distance function (i.e., $d(x)$ in Fig. 5) from the skeleton to the boundary of the component play the essential role. Since MMV is the dual version of the MMC approach, the technique developed in [24] can also be used to construct the TDFs of both star and non-star-shaped voids. On the other hand, if only the TDFs of star-shaped regions are considered, numerous methods can be employed. We refer the readers to [23, 28-31] for the 2D treatments where the TDFs are constructed with use of a center point and a radial distance function associated with the boundary of a star-shaped region. Actually, these treatments can be viewed as the special cases of the treatment applicable for non-star-shaped regions when the skeleton of a non-star-shaped region degenerates to a center point of a star-shaped region and the distance functions (i.e., $d(x)$) are represented in different ways.

For 3D star-shaped regions, following the line of 2D treatment, we propose to use the following method to construct their TDFs. Let $\Omega$ be a closed region which is star-shaped with respect to its central point $\boldsymbol{P}_0=(x_0,y_0,z_0)$, then the TDF of $\Omega$ can be written as

$$\chi(x,y,z;x_0,y_0,z_0)=\sqrt{(x-x_0)^2+(y-y_0)^2+(z-z_0)^2}-r(\theta,\psi), \tag{1}$$

where $r=r(\theta,\psi)$ is the distance from a point $\boldsymbol{P}$ (whose longitude angle and latitude angle with respect to $\boldsymbol{P}_0$ are $\theta$ and $\psi$, respectively) on $\partial\Omega$ to $\boldsymbol{P}_0$ (see Fig. 6 for reference). Accordingly, the domain occupied by the star-shaped void can be identified by $\chi(x,y,z;x_0,y_0,z_0)$ as $\Omega=\{(x,y)|\chi(x,y,z;x_0,y_0,z_0)\le 0\}$. There are actually numerous ways to construct $r=r(\theta,\psi)$. For example, we can use the following Hermitian interpolation scheme to interpolate $r=r(\theta,\psi)$:

$$r(\theta,\psi)=r_1(\theta)r_2(\psi)\big(\psi(\pi-\psi)\big)+c_1\psi^2+c_2(\pi-\psi)^2, \tag{2}$$

where $r_1(\theta)$ and $r_2(\psi)$ are two interpolation functions along the longitude and latitude directions, respectively. In order to guarantee the $C^1$ smoothness of $\partial\Omega$, $r_1(\theta)$, $r_2(\psi)$, $c_1$ and $c_2$ can be interpolated as

$$r_1(\theta)=H_1^{(0)}(\xi)r_i^\theta+H_2^{(0)}(\xi)r_{i+1}^\theta+H_1^{(1)}(\xi)f_i^\theta+H_2^{(1)}(\xi)f_{i+1}^\theta,$$

$$\theta_i\le\theta<\theta_{i+1},\theta_i=\frac{2i\pi}{n},\ i=0,\dots,n-1, \tag{3a}$$

$$r_2(\psi) = H_1^{(0)}(\eta) r_j^{\psi} + H_2^{(0)}(\eta) r_{j+1}^{\psi} + H_1^{(1)}(\eta) f_j^{\psi} + H_2^{(1)}(\eta) f_{j+1}^{\psi},$$

$$\psi_j \le \psi < \psi_{j+1}, \psi_j = \frac{j\pi}{m},\ j = 0, \ldots, m-1, \qquad (3b)$$

$$c_1 = \frac{r_m^{\psi}}{n} \sum_{i=1}^{n} r_i^{\theta}, \qquad (3c)$$

$$c_2 = \frac{r_0^{\psi}}{n} \sum_{i=1}^{n} r_i^{\theta} \qquad (3d)$$

with

$$r_0^{\theta} = r_n^{\theta} \qquad (3e)$$

and

$$f_0^{\theta} = f_n^{\theta}, \qquad (3f)$$

respectively. In Eq. (3a) and Eq. (3b),

$$H_1^{(0)}(\xi) = 1 - 3\xi^2 + 2\xi^3, \qquad (4a)$$

$$H_2^{(0)}(\xi) = 3\xi^2 - 2\xi^3, \qquad (4b)$$

$$H_1^{(1)}(\xi) = \xi - 2\xi^2 + \xi^3, \qquad (4c)$$

$$H_2^{(2)}(\xi) = \xi^3 - \xi^2, \qquad (4d)$$

are basis functions for Hermite interpolation with

$$\xi = \frac{\theta - \theta_i}{\theta_{i+1} - \theta_i}, \qquad (4e)$$

$$\eta = \frac{\psi - \psi_i}{\psi_{i+1} - \psi_i}. \qquad (4f)$$

The values of $r_i^{\theta}, f_i^{\theta}, r_j^{\psi}, f_j^{\psi}, i = 0, \ldots, n; j = 0, \ldots, m$ in Eq. (3) are design variables at each interpolation point. Alternatively, $r(\theta, \psi)$ can also be interpolated as

$$r(\theta, \psi) = \left\| \boldsymbol{S}\left(u(\theta), v(\psi)\right) \right\| = \left\| \sum_{i=0}^{n} \sum_{j=0}^{m} N_{i,p}\big(u(\theta)\big) N_{j,q}\big(v(\psi)\big) \boldsymbol{P}^{i,j} \right\|, \qquad (5)$$

where $N_{i,p}(u)$, $N_{j,q}(v)$ are B-spline basis functions, and $\boldsymbol{P}^{i,j} = \left(P_x^{i,j}, P_y^{i,j}, P_z^{i,j}\right)^{\top}, i = 0, \ldots, n; j = 0, \ldots, m$ are control points with

$$P_x^{i,j} = r^{i,j} \sin(\psi_j) \cos(\theta_i), \qquad (6a)$$

$$P_y^{i,j} = r^{i,j} \sin(\psi_j) \sin(\theta_i), \qquad (6b)$$

$$P_z^{i,j} = r^{i,j}\cos(\psi_j), \tag{6c}$$

$$\theta_i = \frac{2i\pi}{n}, \quad i = 0, \dots, n \tag{6d}$$

and

$$\psi_j = \frac{j\pi}{m}, \quad j = 0, \dots, m, \tag{6e}$$

respectively. With use of this interpolation scheme, $r^{i,j}$, $i = 0, \dots, n$; $j = 0, \dots, m$ are design variables. Since $\Omega$ is a closed region, it is also required that

$$r^{0,j} = r^{n,j}, \quad j = 0, \dots, m, \tag{7a}$$

$$r^{i,0} = r^{0,0}, \quad i = 1, \dots, n, \tag{7b}$$

$$r^{i,m} = r^{0,m}, \quad i = 1, \dots, n \tag{7c}$$

in this scheme.

The advantage of the tensor product interpolation scheme in Eq. (2) is that the total number of its design variables is usually less than that of the interpolation scheme in Eq. (5). The latter, however, is more flexible in realizing a local control of the shape of the boundary surface (i.e., $\partial\Omega$).

In summary, under the above geometry description, the problem formulation of a 3D TopOpt problem under the proposed MMV-based framework can be expressed as

$$\text{Find} \quad \boldsymbol{D} = \left((\boldsymbol{D}^1)^\top, \dots, (\boldsymbol{D}^i)^\top, \dots, (\boldsymbol{D}^{nv})^\top\right)^\top, \ \boldsymbol{u}(\boldsymbol{x})$$

$$\text{Minimize} \ I = I(\boldsymbol{D})$$

s.t.

$$\int_{\mathrm{D}} H(\chi^s(\boldsymbol{x};\boldsymbol{D}))\boldsymbol{\varepsilon}(\boldsymbol{u}):\mathbb{E}^s:\boldsymbol{\varepsilon}(\boldsymbol{v})\mathrm{dV} = \int_{\mathrm{D}} H(\chi^s(\boldsymbol{x};\boldsymbol{D}))\boldsymbol{f}\cdot\boldsymbol{v}\mathrm{dV} + \int_{\Gamma_{\mathrm{t}}} \boldsymbol{t}\cdot\boldsymbol{v}\mathrm{dS},$$

$$\forall \boldsymbol{v} \in \mathcal{U}_{\mathrm{ad}},$$

$$\int_{\mathrm{D}} H(\chi^s(\boldsymbol{x};\boldsymbol{D}))\mathrm{dV} \le \bar{V},$$

$$g_j(\chi^s(\boldsymbol{x};\boldsymbol{D})) \le 0, \ j = 1, \dots, m,$$

$$\boldsymbol{D} \subset \mathcal{U}_{\boldsymbol{D}},$$

$$\boldsymbol{u} = \bar{\boldsymbol{u}}, \quad \text{on } \Gamma_{\mathrm{u}}, \tag{8}$$

where $\boldsymbol{D}^i = \left(x_0^i, y_0^i, z_0^i, r_i^{0,0}, \dots, r_i^{n,m}\right)^\top$ (in the NURBS interpolation scheme) or $\boldsymbol{D}^i = \left(x_0^i, y_0^i, z_0^i, r_{0i}^\theta, f_{0i}^\theta, \dots, r_{ni}^\theta, f_{ni}^\theta, r_{0i}^\psi, f_{0i}^\psi, \dots, r_{mi}^\psi, f_{mi}^\psi\right)^\top$ (in the Hermite interpolation scheme) , $i =$ $1,\dots,nv$ is the vector containing the design variables associated with the boundary surface of the $i$-th void; $nv$ is the total number of voids in the design domain; $\boldsymbol{f}$ denotes the body force density in $\Omega^s$ and $\boldsymbol{t}$ is the surface traction on Neumann boundary $\Gamma_t$; $\mathbb{E}^s = E^s/(1+\nu^s)[\mathbb{I} + \nu^s/(1-2\nu^s)\,\boldsymbol{\delta}\otimes\boldsymbol{\delta}]$ ($\mathbb{I}$ and $\boldsymbol{\delta}$ denote the fourth and the second order identity tensor, respectively) is the fourth order isotropic elasticity tensor of the solid material constituting the components with $E^s$ and $\nu^s$ denoting the Young's modulus and Poisson's ratio of the solid material, respectively; $\chi^s$ is the TDF of the whole structure which can be constructed from the TDF of each void (i.e., $\chi_i = \chi_i(\boldsymbol{x}),\ i = 1,\dots,nv$) as $\chi^s = \min(\chi_1,\dots,\chi_n)$; $H = H(x)$ is the Heaviside function defined as $H(x) = 1, \text{if } x \geq 0;\ H(x) = 1.0\times10^{-3}, \text{otherwise}$; $\boldsymbol{u}$ and $\boldsymbol{v}$ are the displacement field and the test function, respectively; $\bar{\boldsymbol{u}}$ is the prescribed displacement on Dirichlet boundary $\Gamma_u$ and $\mathcal{U}_{ad} = \{\boldsymbol{v} |\ \boldsymbol{v} \in \mathbf{H}^1(\Omega), \boldsymbol{v} = \mathbf{0} \text{ on } \Gamma_u\}$ is the admissible set of the test function. In Eq. (8), $\mathcal{U}_{\boldsymbol{D}}$ denote the feasible set of $\boldsymbol{D}$. Furthermore, the symbol $\bar{V}$ is the upper bound of the available volume of solid material. $g_j$, $j = 1,\dots,m$ are some other concerned constraint functions/functionals in the optimization problem.

## 4. Numerical solution aspects

In this section, we shall discuss the relevant issues for the numerical implementation of the proposed approach.

### 4.1 Finite element analysis

In the present study, the fixed mesh FE method proposed in [32] is adopted to calculate the structural response without re-meshing. This method is constructed based on the well-known ersatz material model [11, 12] of topology optimization. As pointed out in [20, 32], in this approach, the stiffness matrix of the $e$-th finite element is calculated as

$$\mathbf{K}_e = \int_{\Omega_e} \mathbf{B}^{\top}\mathbf{D}_e^*\mathbf{B}\,\mathrm{dV}, \tag{9}$$

where $\mathbf{B}$ is the strain matrix and $\Omega_e$ represents the region occupied by the $e$-th finite element. In Eq. (9), $\mathbf{D}_e^* = \rho_e\mathbf{D}_0$ with $\mathbf{D}_0$ and $\rho_e$ denoting the elasticity matrix corresponding to the solid material and the volume fraction of solid material in $\Omega_e$, respectively. In the present work, the volume fraction $\rho_e$ of the element is interpolated as

$$\rho_e = \frac{\sum_{i=1}^{8} H_\epsilon\left((\chi^s)_i^e\right)}{8}, \tag{10}$$

where $(\chi^s)_i^e,\ i=1,\dots,8$ are the values of the TDF $\chi^s(\boldsymbol{x})$ on each node of $\Omega_e$. In Eq. (10), $H_\epsilon(x)$ is the regularized Heaviside function which can be expressed as

$$H_\epsilon(x) = \begin{cases} 1, & \text{if } x > \epsilon, \\ \dfrac{3(1-\gamma)}{4}\left(\dfrac{x}{\epsilon} - \dfrac{x^3}{3\epsilon^3}\right) + \dfrac{(1+\gamma)}{2}, & \text{if } -\epsilon \le x \le \epsilon, \\ \gamma, & \text{otherwise.} \end{cases} \tag{11}$$

In Eq. (11), $\epsilon$ describes the width of numerical approximation and $\gamma$ is a small positive number to ensure that the global stiffness matrix of the structure is nonsingular.

In traditional material density based topology optimization approach, the finite element analysis is usually carried out on the entire design domain ($\mathrm{i.e.,D}$). This is actually due to two reasons. The first is that the topology change of the structure is achieved by emerging/deleting of material in the design domain and the second is that the optimization model and the analysis model are strongly coupled in this kind of approach. In the present explicit boundary evolution-based approach, the computational effort associated with FEA can be reduced substantially by removing the unnecessary DOFs from the FE model at every step of numerical optimization. This can be achieved as follows. Firstly, the global stiffness matrix $\mathbf{K}$ corresponding to the entire design domain $\mathrm{D}$ is assembled in a straightforward way. Then loop for every node of the fixed FE mesh discretized the design domain. If all the neighboring elements sharing one specific node (e.g., node $i$) are constituted by pure weak material (i.e., $(\chi^s)_j^e < 0,\ j = 1,\dots,8$), then label this node as $T_i = -1$. Finally deleting all DOFs from $\mathbf{K}$ associated with the nodes whose $T$ value are equal to $-1$ and forming a new global stiffness matrix $\widetilde{\mathbf{K}}$ with reduced number of dimension. We refer the readers to Fig. 7 for a schematic illustration of the above operations. The modified matrix $\widetilde{\mathbf{K}}$ will then be used for subsequent FEA and the displacement vector $\widetilde{\boldsymbol{u}}$ obtained from $\widetilde{\boldsymbol{u}} = \widetilde{\mathbf{K}}^{-1}\tilde{\boldsymbol{f}}$ (here $\tilde{\boldsymbol{f}}$ is the corresponding modified external load vector) will be used for calculating the sensitivity.

Since in the proposed MMV-based approach, numerical optimization is always started from a feasible design containing a reasonable load transferring path and the optimizer will try to improve the value of objective function and at the same time respect the prescribed constraints, then it is highly possible (and has been verified by numerous numerical experiments) that there always exists a load transferring path in solid region during the whole process of optimization if the move limits of design variables are not too large. Therefore the above treatment is also applicable even though some small isolated islands may appear in the design domain when structural topology is evolved. This fact has also been verified by the numerical examples provide in the next section. With use of the above treatment, the time for FEA can be reduced substantially (e.g., 10 times) especially when the allowable volume of solid material is small. Compared to the DOF removal technique developed for variable density approach in [33], the treatment described above is more straightforward and can be implemented in a numerically robust way. It is also worthy of noting that although the aforementioned DOF removal technique is developed for the MMV-based approach, it is actually also applicable to the MMC-based approach. The essential difference is that in the MMC-based approach the removal of DOFs is only allowed when all components have been connected and formed a load transferring path in the design domain (see Fig. 8 for reference). Corresponding technical details on this aspect will be reported in a separate work.

### 4.2 Sensitivity analysis

The sensitivity analysis can be performed in a similar way as in [32]. For a general objective/constraint functional $I$ under consideration, generally the sensitivity of $I$ with respect to a design variable $a$ can be expressed in a discretized adjoint approach setting as

$$\frac{\partial I}{\partial a} = \widetilde{\boldsymbol{w}}^{\top} \frac{\partial \widetilde{\mathbf{K}}}{\partial a} \widetilde{\boldsymbol{u}} = \widetilde{\boldsymbol{w}}^{\top} \frac{E^s}{8} \sum_{e=1}^{NE} \sum_{j=1}^{8} \left( \frac{\partial H_\epsilon(x)}{\partial x} \Big| x = (\chi^s)_j^e \right) \frac{\partial \left( (\chi^s)_j^e \right)}{\partial a} \boldsymbol{k_s}\, \widetilde{\boldsymbol{u}}, \tag{12}$$

where $\widetilde{\boldsymbol{w}} = \widetilde{\mathbf{K}}^{-\mathbf{1}} \left( \frac{\partial I}{\partial \widetilde{\boldsymbol{u}}} \right)$ is the adjoint displacement vector and $\boldsymbol{u}$ is the primary displacement vector, respectively. In Eq. (12), $\widetilde{\mathbf{K}}$ is the modified global stiffness matrix of the structure and $\boldsymbol{k_s}$ is the element stiffness matrix corresponding to $(\chi^s)_j^e = 1, j = 1, \ldots, 8$ and the Young's modulus of the solid material $E^s = 1$. The symbol $NE$ in Eq. (12) denotes the total number of elements in the design domain. From numerical implementation point of view, the key point for sensitivity analysis under the above treatment is to calculate the quantity of $\partial \chi^s / \partial a$. As pointed out in [21, 32], this can be achieved easily in the proposed approach since $\chi^s$ is an explicit function of $a$. Detailed expressions of $\partial \chi^s / \partial a$ under different geometry description schemes are listed in the Appendix. Furthermore, as shown in [32] $\partial \chi^s / \partial a$ can also be obtained approximately by the finite difference quotient approach with enough accuracy.

## 5. Numerical example

In this section, several numerical examples are provided to demonstrate the effectiveness of the proposed approach for designing optimal 3D structures. The considered objective and constraint functionals are the structural compliance and the available volume of the solid material, respectively. Since the main purpose of the present section is to test the numerical performance of the suggested approach, the material, load and geometric data are all chosen as dimensionless unless otherwise stated. The Young's modulus and Poisson's ratio of the solid material are taken as $E^s = 1$ and $\nu^s = 0.3$, respectively. Uniform eight-node hexahedra (8-node) elements are used for FE discretization in all examples examined. The optimization problems are solved by the well-known Method of Moving Asymptotes (MMA) method [34]. Furthermore, all computations are carried out on Dell-T5810 (3.7GHz with 128GB memory).

### 5.1 Short cantilever beam example

Firstly, the classical short cantilever beam problem is examined. The geometry, boundary condition, and external load of the considered problem is shown in Fig. 9. A $10 \times 2 \times 5$ cuboid design domain is discretized by a $40 \times 10 \times 20$ FE mesh. The left side of the design domain is fixed and distributed vertical loads with magnitude of 1 are imposed on the bottom of the right side of the design domain. The upper bound of available solid material volume is set to $\bar{V} = 0.15 \times |\mathrm{D}| = 15$.

Two kinds of initial designs composed of 34 voids is shown in Fig. 10 and Hermite and NURBS interpolation scheme are adopted to describe these voids, respectively. The final optimized structures are shown in Fig. 11 and the corresponding values of the objective functional are $I = 127.10$ and $I = 128.15$, which are almost the same for these two interpolation schemes. From the intermediate steps of the optimization process shown in Fig. 12, it can be observed that the optimal structural topology has been achieved gradually through the morphing, overlapping and hiding of the voids. Compared with the level set method where optimal structural topology is also obtained from a set of voids distributed in the design domain, the distinctive feature of the MMV approach is that the evolution of the structural boundary is driven explicitly by a set of parameters which have clear geometrical meanings.

It is worthwhile to note that there are only $1394$ (for Hermite interpolation) and $2210$ (for NURBS interpolation) design variables in the proposed problem, which is totally independent of the resolution of the finite element mesh and much less than those employed in traditional methods (about $40 \times 10 \times 20 = 8000$ for the adopted FE resolution). When mathematical programming methods are adopted as the optimizer (such as the MMA used in the proposed work), having relative low number of design variables obviously can reduce the computational time in optimization stage effectively. Furthermore, the histories of the CPU time for FEA in each iteration step is shown in Fig. 13. The point worth noting is that with use of the DOF removal technique suggested in subsection 4.1, the computational effort associated with FEA is reduced substantially as the actual volume of solid material decreases. It can be observed from Fig. 13 that the CPU time for FEA can be reduced from about 1.70 seconds (corresponding to FEA with use of the total number of DOFs) to 0.71 seconds in 200 iteration steps. Therefore the smaller the solid material volume, the faster the FEA in an iteration step. This computational advantage, however, cannot be achieved easily with use of the traditional approaches (especially the element-based ones).

Furthermore, the same problem is also solved by MMV approach with use of a different initial design including more voids. This time, there are totally 55 voids contained in the initial design (as shown in Fig. 14) and each void is still described by the NURBS interpolation scheme. Under this circumstance, the total number of design variables is 3575. The optimized structure is shown in Fig. 15 and the corresponding value of the objective functional is $I = 128.76$. Since the available volume fraction of the solid material does not change, the CPU time for FEA in this case is almost the same as that in the above case (with 34 voids in the initial design).

**5.2 L-shape chair example**

In this example, an L-shape chair problem is considered. The design domain, external load, and boundary conditions are all shown in Fig. 16. A $4 \times 3 \times 4$ cuboid zone on the left top of the domain is set as a non-designable void domain. Two distributed loads are imposed on the surface of non-designable solid domain (highlighted in deep color) by applying the loads with density of 1 and 0.2, respectively. The design domain is discretized by a $60 \times 30 \times 60$ FE mesh and the upper bound of the available solid material for this problem is set to $\bar{V} = 0.1 \times |\mathrm{D}| = 10.8$.

For this example, only NURBS interpolation is used for describing the shape of each void. There are $66$ voids in the initial design as shown in Fig. 17 and the total number of design variables is $66 \times 65 = 4290$. The corresponding optimized structure with an objective functional value of $3266.20$ is shown in Fig. 18. From Fig. 18, it can be seen that an effective load transmission path is constructed with a set of voids with complex shapes. Two curved and four straight load transmission paths are obtained on the back and bottom of the non-designable solid domain to support the structure, respectively. Fig. 18 also plots a specific void that constitutes the no material part of the optimized structure (indicated in green). In this part, the void is enclosed by a single NURBS surface with only 31 active interpolation points. This demonstrates clearly the capability of the proposed geometry representation scheme to describe structural boundary with a complex shape. The history of the CPU time for FEA is shown in Fig. 19. In the early stage of optimization (before 200 step), the FEA is carried out with use of almost the total number of DOFs (about 346000) and the corresponding CPU time is about 180.00 seconds. After 200 step, however, the FEA is typically performed with use of a reduced number of DOFs (about 87000) and the corresponding CPU time is reduced to 11.10 seconds, which is only $1/16$ of that without using the DOF removal technique. It is reasonable to be expected that the saving of CPU time can be even higher when more elements are used for FE discretization. Iteration histories of the structural compliance and volume during the course of optimization are also shown in Fig. 20.

### 5.3 Torsion beam example

In order to further demonstrate the flexibility of the proposed approach to deal with 3D topological changes, the torsion beam example shown in Fig. 21 is also considered. The geometry of the design domain, the boundary condition, and the external load are all depicted in Fig. 21. For this example, the $12 \times 4 \times 4$ cuboid design domain is discretized by a $96 \times 32 \times 32$ FE mesh. The left and right sides (both have a dimension of $0.25 \times 4 \times 4$) are defined as non-designable solid domain. Four loads are imposed on four vertices of the right side of the design domain, respectively. The constraint of available solid material volume is taken as $V \leq 0.15 \times |\mathrm{D}| = 28.8$.

As shown in Fig. 22a, the initial design of the problem consists of 106 voids, and each void is described by a NURBS surface with 62 interpolation points and one center point. This means that the total number of design variables is $(3 + 62) \times 106 = 6890$, which is much smaller than that associated with the traditional approaches under the same FE mesh (i.e., $96 \times 32 \times 32 = 98304$).

Fig. 22b plots the corresponding optimized structure obtained through the proposed approach. It can be observed from this figure that the optimized structure is actually constituted by a cylinder shell-like network of solid bars with small thicknesses. By comparing this optimized form of material distribution with the initial one, the flexibility of the proposed geometry representation scheme is clearly demonstrated. The same problem is also solved with use of the MMC-based approach [20] from the initial design shown in Fig. 23a. This initial design is composed of 128 components and therefore the total number of design variables is only $128 \times 9 = 1152$. Fig. 23b plots the optimized structure obtained with use of the MMC approach. Fig. 22b and Fig. 23b indicate that the optimized structures obtained by the MMV and MMC-based approaches are similar in nature in the sense that both of them are cylinder shell-like lattice structures and one structure can be viewed as being produced by a rigid body translation of another one along the axial direction. The main difference between the two optimized structures is that the load transmission path is directly connected with loading points in the MMV-based result while in the MMC-based result, the loading path is originated from the middle points of the right side of the non-designable domain. The values of the structural compliance associated with these two optimized structures are 2643.54 (MMV-based approach) and 2742.12 (MMC-based approach), respectively. The histories of the CPU time for FEA are also shown in Fig. 24. This figure demonstrates once again that the proposed DOF removal technique is very effective to enhance the solution efficiency of 3D TopOpt problems under the MMV/MMC-based topology optimization framework. As the optimization iteration proceeds, the solution time of FEA can be reduced by nearly two orders of magnitude in both MMV and MMC approaches since the DOFs involved in FEA have been reduced from $3 \times 10^5$ at the beginning of optimization to $1 \times 10^5$ after 100 (in the MMV approach) or 20 (in the MMC approach) iteration steps. Some intermediate results obtained during the courses of optimization associated with the MMV-based and MMC-based are also provided in Fig. 25, respectively. It can be seen from these figures that MMC-based approach is more efficient to construct an effective load transmission path (only taking about 10 iteration steps) while the convergence is more stable for the MMV-based approach. Finally, it is also worth noting that all optimized structures obtained the MMV/MMC-based approaches are actually pure black-and-white and contain no grey elements which are unavoidable in traditional approaches especially for 3D problems.

It is worth noting that it is not appropriate to make a direct comparison of the total solution times associated with different methods. This is because the total solution time is dependent on the total iteration numbers which is largely determined by numerical implementation details (e.g., the size of filter radius, the parameter for continuation, the parameters in MMA optimizer, etc.), and most important of all, we cannot guarantee that we have implemented the methods for comparison in the most efficient way. Under this circumstance, it seems reasonable to provide the solution times corresponding to different stages of computation in one representative iteration step of different approaches to demonstrate the performances of our method. For this example, it is found that at step 2, the solution times (second) of FEA in MMV and SIMP approaches (using the open source code provided in [35]) are 109.42 and 116.86, respectively; the solution times of MMA in MMV and SIMP approaches are 0.06 and 47.61, respectively. At step 150, the solution times of FEA in MMV and SIMP approaches are 20.36 and 106.95, respectively; the solution times of MMA in MMV and SIMP approaches are 0.06 and 46.04, respectively.

## 6. Concluding remarks

In the present paper, a new approach for solving 3D TopOpt problems is developed based on the Moving Morphable Void (MMV) approach first initialized in [22, 23]. The most attractive feature of this approach is that it can not only reduce the total number of design variables associated with the optimization problem but also provide the possibility for cutting down the number of DOFs substantially through an effective element removal technique. The later treatment is achievable since in the proposed solution approach the optimization and analysis models are totally decoupled. Numerical examples indicate that the two challenging issues related to 3D TopOpt problems, namely, the large numbers of design variables and DOFs, have been well addressed with use of the proposed approach. Besides, as the same in the MMC approach, the resolution of the shape and topology of an optimized structure is totally independent on the mesh resolution for FEA. Another interesting research topic is using boundary element method (BEM) for structural response analysis. On this aspect, the conventional BEM and the recently developed isogeometric BEM method [36] can all be employed to reduce the DOFs for structural analysis. It is also worth noting that the present approach is only implemented in a single core computing environment. It is reasonable to expect that much more solution efficiency can be obtained if the proposed approach is implemented in a parallel computing environment. Research work along this direction is currently being carried out intensively.

## Appendix

Under the Hermite geometry description scheme, the expression of the derivative of $\chi$ with respect to each design variable in Eq. (12) can be calculated as follows:

$$\frac{\partial \chi}{\partial x_0} = \frac{x_0}{\sqrt{(x-x_0)^2+(y-y_0)^2+(z-z_0)^2}}, \tag{A1a}$$

$$\frac{\partial \chi}{\partial y_0} = \frac{y_0}{\sqrt{(x-x_0)^2+(y-y_0)^2+(z-z_0)^2}}, \tag{A1b}$$

$$\frac{\partial \chi}{\partial z_0} = \frac{z_0}{\sqrt{(x-x_0)^2+(y-y_0)^2+(z-z_0)^2}}, \tag{A1c}$$

$$\frac{\partial \chi}{\partial r_i^\theta} = \frac{\partial r(\theta,\psi)}{\partial r_i^\theta} = \frac{\partial r_1(\theta)}{\partial r_i^\theta} r_2(\psi)\big(\psi(\pi-\psi)\big) + \frac{r_m^\psi}{n}\psi^2 + \frac{r_0^\psi}{n}(\pi-\psi)^2, \tag{A1d}$$

$$\frac{\partial \chi}{\partial f_i^\theta} = \frac{\partial r(\theta,\psi)}{\partial f_i^\theta} = \frac{\partial r_1(\theta)}{\partial f_i^\theta} r_2(\psi)\big(\psi(\pi-\psi)\big), \tag{A1e}$$

$$\frac{\partial \chi}{\partial r_j^\psi} = \frac{\partial r(\theta,\psi)}{\partial r_j^\psi} = \begin{cases} r_1(\theta)\dfrac{\partial r_2(\psi)}{\partial r_j^\psi}\big(\psi(\pi-\psi)\big) + \dfrac{\psi^2}{n}\displaystyle\sum_{i=1}^{n} r_i^\theta, & \text{if } j=0, \\ r_1(\theta)\dfrac{\partial r_2(\psi)}{\partial r_j^\psi}\big(\psi(\pi-\psi)\big) + \dfrac{(\pi-\psi)^2}{n}\displaystyle\sum_{i=1}^{n} r_i^\theta, & \text{if } j=m, \\ r_1(\theta)\dfrac{\partial r_2(\psi)}{\partial r_j^\psi}\big(\psi(\pi-\psi)\big), & \text{otherwise,} \end{cases} \tag{A1$f$}$$

$$\frac{\partial \chi}{\partial f_j^\psi} = \frac{\partial r(\theta,\psi)}{\partial f_j^\psi} = r_1(\theta)\frac{\partial r_2(\psi)}{\partial f_j^\psi}\big(\psi(\pi-\psi)\big), \tag{A1g}$$

where

$$\frac{\partial r_1(\theta)}{\partial r_i^\theta} = \begin{cases} H_2^{(0)}(\xi), & \text{if } \theta_{i-1}\le\theta<\theta_i, i=1,\dots,n, \\ H_1^{(0)}(\xi), & \text{if } \theta_i\le\theta<\theta_{i+1}, i=0,\dots,n-1, \\ 0, & \text{otherwise,} \end{cases} \tag{A2a}$$

$$\frac{\partial r_1(\theta)}{\partial f_i^\theta} = \begin{cases} H_2^{(1)}(\xi), & \text{if } \theta_{i-1}\le\theta<\theta_i, i=1,\dots,n, \\ H_1^{(1)}(\xi), & \text{if } \theta_i\le\theta<\theta_{i+1}, i=0,\dots,n-1, \\ 0, & \text{otherwise,} \end{cases} \tag{A2b}$$

$$\frac{\partial r_2(\psi)}{\partial r_j^\psi} = \begin{cases} H_2^{(0)}(\eta), & \text{if } \psi_{j-1}\le\psi<\psi_j, j=1,\dots,m, \\ H_1^{(0)}(\eta), & \text{if } \psi_j\le\psi<\psi_{j+1}, j=0,\dots,m-1, \\ 0, & \text{otherwise,} \end{cases} \tag{A2c}$$

$$\frac{\partial r_2(\psi)}{\partial f_j^{\psi}} = \begin{cases} H_2^{(1)}(\eta), & \text{if } \psi_{j-1} \le \psi < \psi_j, j = 1, \dots, m, \\ H_1^{(1)}(\eta), & \text{if } \psi_j \le \psi < \psi_{j+1}, j = 0, \dots, m-1, \\ 0, & \text{otherwise.} \end{cases} \quad \text{(A2d)}$$

Under the NURBS geometry description scheme, the corresponding expressions are

$$\frac{\partial \chi}{\partial x_0} = \frac{x_0}{\sqrt{(x-x_0)^2 + (y-y_0)^2 + (z-z_0)^2}}, \quad \text{(A3a)}$$

$$\frac{\partial \chi}{\partial y_0} = \frac{y_0}{\sqrt{(x-x_0)^2 + (y-y_0)^2 + (z-z_0)^2}}, \quad \text{(A3b)}$$

$$\frac{\partial \chi}{\partial z_0} = \frac{z_0}{\sqrt{(x-x_0)^2 + (y-y_0)^2 + (z-z_0)^2}}, \quad \text{(A3c)}$$

$$\frac{\partial \chi}{\partial r^{i,j}} = \frac{\partial r(\theta,\psi)}{\partial r^{i,j}}, i = 0, \dots, n, j = 0, \dots, m, \quad \text{(A3d)}$$

$$\frac{\partial r(\theta,\psi)}{\partial r^{i,j}} = \frac{1}{r(\theta,\psi)} \left( S_x(\theta,\psi) \frac{\partial S_x(\theta,\psi)}{\partial r^{i,j}} + S_y(\theta,\psi) \frac{\partial S_y(\theta,\psi)}{\partial r^{i,j}} + S_z(\theta,\psi) \frac{\partial S_z(\theta,\psi)}{\partial r^{i,j}} \right),$$

$$i = 0, \dots, n, j = 0, \dots, m, \text{(A3e)},$$

respectively and

$$\frac{\partial S_x(u,v)}{\partial r^{k,l}} = \sum_{i=0}^{n} \sum_{j=0}^{m} N_{i,p}(u) N_{j,q}(v) \frac{\partial P_x^{i,j}}{\partial r^{k,l}}, i = 0, \dots, n, j = 0, \dots, m, \quad \text{(A4a)}$$

$$\frac{\partial S_y(u,v)}{\partial r^{k,l}} = \sum_{i=0}^{n} \sum_{j=0}^{m} N_{i,p}(u) N_{j,q}(v) \frac{\partial P_y^{i,j}}{\partial r^{k,l}}, i = 0, \dots, n, j = 0, \dots, m, \quad \text{(A4b)}$$

$$\frac{\partial S_z(u,v)}{\partial r^{k,l}} = \sum_{i=0}^{n} \sum_{j=0}^{m} N_{i,p}(u) N_{j,q}(v) \frac{\partial P_z^{i,j}}{\partial r^{k,l}}, i = 0, \dots, n, j = 0, \dots, m. \quad \text{(A4c)}$$

## Acknowledgements

The financial supports from the National Key Research and Development Plan (2016YFB0201600), the National Natural Science Foundation (11402048, 11372004), Program for Changjiang Scholars, Innovative Research Team in University (PCSIRT) and 111 Project (B14013) are gratefully acknowledged.

## References

[1] W. Prager, G.I.N. Rozvany, Optimal layout of grillages, J. Struct. Mech., 5 (1) (1977) 1-18.

[2] K.T. Cheng, N. Olhoff, An investigation concerning optimal design of solid elastic plate, Int. J. Solids Struct., 17 (3) (1981) 305-323.

[3] M.P. Bendsøe, N. Kikuchi, Generating optimal topologies in structural design using a homogenization method, Comput. Methods Appl. Mech. Engrg., 71 (2) (1988) 197-224.

[4] M. Zhou, G.I.N. Rozvany, The COC algorithm, Part II: topological, geometrical and generalized shape optimization, Comput. Methods Appl. Mech. Engrg., 89 (1-3) (1991) 309-336.

[5] G.I.N. Rozvany, A critical review of established methods of structural topology optimization, Struct. Multidiscip. Optim., 37 (3) (2009) 217-237.

[6] X. Guo, G.D. Cheng, Recent development in structural design and optimization, Acta Mech. Sin., 26 (6) (2010) 807-823.

[7] O. Sigmund, K. Maute, Topology optimization approaches A comparative review, Struct. Multidiscip. Optim., 48 (6) (2013) 1031-1055.

[8] N.P. van Dijk, K. Maute, M. Langelaar, F. van Keulen, Level-set methods for structural topology optimization: a review, Struct. Multidiscip. Optim., 48 (3) (2013) 437-472.

[9] M.P. Bendsøe, Optimal shape design as a material distribution problem, Struct. Optim., 1 (4) (1989) 193-202.

[10] J.D. Deaton, R.V. Grandhi, A survey of structural and multidisciplinary continuum topology optimization: post 2000, Struct. Multidiscip. Optim., 49 (1) (2014) 1-38.

[11] M.Y. Wang, X.M. Wang, D.M. Guo, A level set method for structural topology optimization, Comput. Methods Appl. Mech. Engrg., 192 (1-2) (2003) 227-246.

[12] G. Allaire, F. Jouve, A.M. Toader, Structural optimization using sensitivity analysis and a level-set method, J. Comput. Phys., 194 (1) (2004) 363-393.

[13] C. Fleury, Structural optimization methods for large scale problems: Status and limitations, in: ASME International Design Engineering Technical Conferences/ Computers and Information in Engineering Conference, Las Vegas, NV, 2007, pp. 513-522.

[14] T. Zegard, G.H. Paulino, Bridging topology optimization and additive manufacturing, Struct. Multidiscip. Optim., 53 (1) (2016) 175-192.

[15] A. Diaz, R. Lipton, Optimal material layout for 3D elastic structures, Struct. Optim., 13 (1) (1997) 60-64.

[16] P. Fernandes, J.M. Guedes, H. Rodrigues, Topology optimization of three-dimensional linear elastic structures with a constraint on "perimeter", Comput. Mech., 73 (6) (1999) 583-594.

[17] C.H. Villanueva, K. Maute, Density and level set-XFEM schemes for topology optimization of 3-D structures, Comput. Mech., 54 (1) (2014) 133-150.

[18] N. Aage, B.S. Lazarov, Parallel framework for topology optimization using the method of moving asymptotes, Struct. Multidiscip. Optim., 47 (4) (2013) 493-505.

[19] K. Liu, A. Tovar, An efficient 3D topology optimization code written in Matlab, Struct. Multidiscip. Optim., 50 (6) (2014) 1175-1196.

[20] W.S. Zhang, D. Li, J. Yuan, J.F. Song, X. Guo, A new three-dimensional topology optimization method based on moving morphable components (MMCs), Comput. Mech., (2016) 1-19. doi:10.1007/s00466-016-1365-0.

[21] X. Guo, W.S. Zhang, W.L. Zhong, Doing Topology Optimization Explicitly and Geometrically—A New Moving Morphable Components Based Framework, J. Appl. Mech., 81 (8) (2014) 081009.

[22] W.Y. Yang, W.S. Zhang, X. Guo, Explicit structural topology optimization via Moving Morphable Voids (MMV) approach, in: 2016 Asian Congress of Structural and Multidisciplinary Optimization, Nagasaki, Japan, 2016, pp. 98.

[23] W.S. Zhang, W.Y. Yang, J.H. Zhou, D. Li, X. Guo, Structural topology optimization through explicit boundary evolution, J. Appl. Mech., 84 (1) (2017) 011011.

[24] X. Guo, W.S. Zhang, J. Zhang, J. Yuan, Explicit structural topology optimization based on moving morphable components (MMC) with curved skeletons, Comput. Methods Appl. Mech. Engrg., 310 (2016) 711-748.

[25] D. Terzopoulos, A. Witkin, M. Kass, Symmetry-seeking models and 3D object reconstruction, Int. J. Comput. Vision, 1 (3) (1987) 211-221.

[26] R.J. Campbell, P.J. Flynn, A survey of free-form object representation and recognition techniques, Comput. Vision Image Understanding, 81 (2) (2001) 166-210.

[27] G. Taubin, Estimation of planar curves, surfaces, and nonplanar space curves defined by implicit equations with applications to edge and range image segmentation, IEEE Trans. Pattern Anal. Mach. Intell., 13 (11) (1991) 1115-1138.

[28] W.H. Zhang, L.Y. Zhao, T. Gao, S.Y. Cai, Topology optimization with closed B-splines and Boolean operations, Comput. Methods Appl. Mech. Engrg., 315 (2017) 652-670.

[29] M.D. Buhmann, Radial basis functions: theory and implementations, Camb. Monogr. Appl. Comput. Math., 12 (2003) 147-165.

[30] E.J. Kansa, H. Power, G.E. Fasshauer, L. Ling, A volumetric integral radial basis function method for time-dependent partial differential equations. I. Formulation, Eng. Anal. Bound. Elem., 28 (10) (2004) 1191-1206.

[31] R.L. Hardy, Multiquadric equations of topography and other irregular surface, J. Geophys. Res., 76 (8) (1971) 1905-1915.

[32] W.S. Zhang, J. Yuan, J. Zhang, X. Guo, A new topology optimization approach based on Moving Morphable Components (MMC) and the ersatz material model, Struct. Multidiscip. Optim., 53 (6) (2016) 1243-1260.

[33] T.E. Bruns, D.A. Tortorelli, An element removal and reintroduction strategy for the topology optimization of structures and compliant mechanisms, Int. J. Numer. Meth. Engrg, 57 (10) (2003) 1413-1430.

[34] K. Svanberg, The method of moving asymptotes—a new method for structural optimization, Int. J. Numer. Meth. Engrg, 24 (2) (1987) 359-373.

[35] K. Liu, A. Tovar, An efficient 3D topology optimization code written in Matlab. Struct. Multidiscip. Optim., 50(6) (2014) 1175–1196.

[36] H. Lian, P. Kerfriden, S.P.A. Bordas, Shape optimization directly from CAD: An isogeometric boundary element approach using T-splines, Comput. Methods Appl. Mech. Engrg., 317 (15) (2017) 1-41.

# Figs

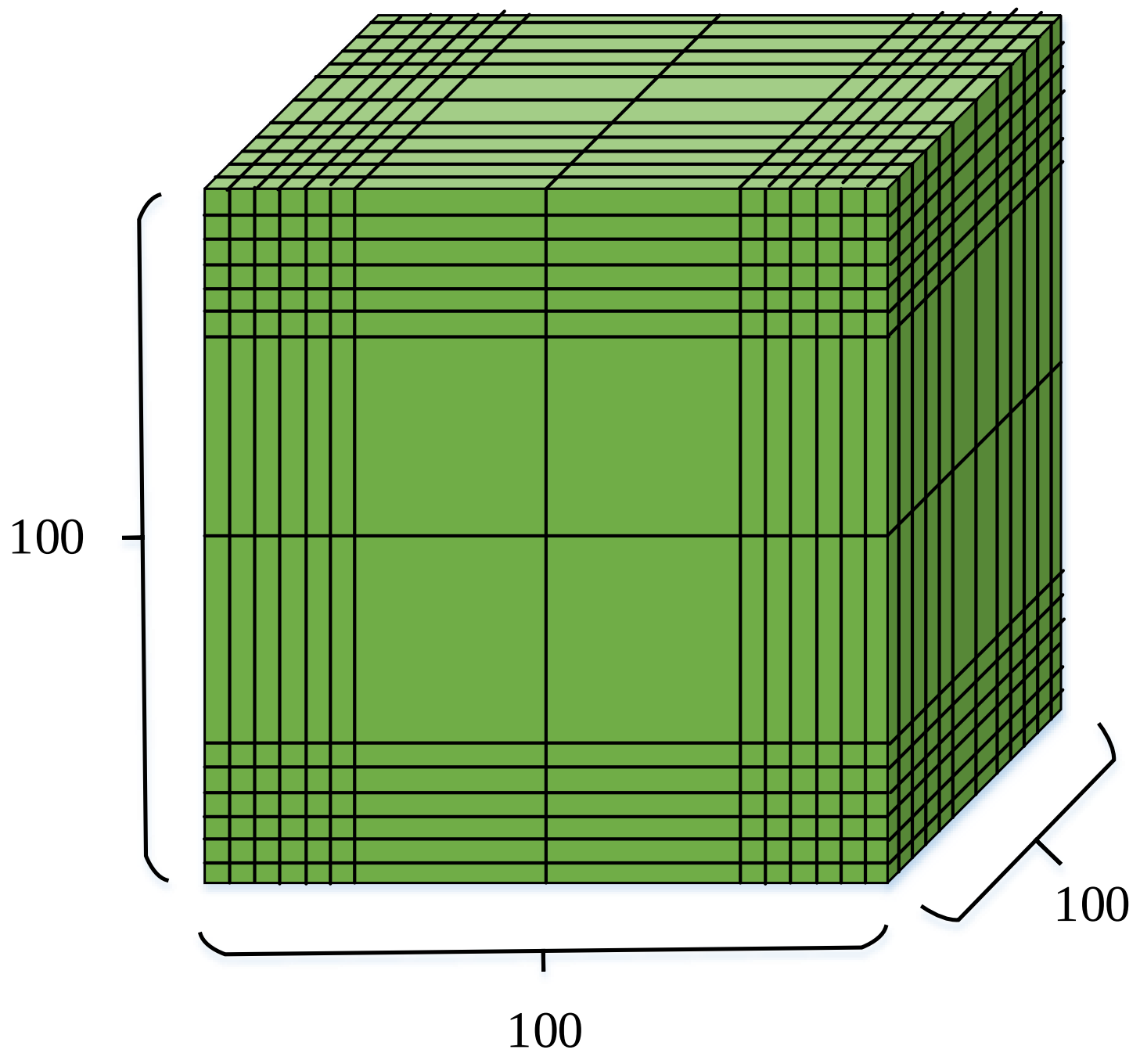

Fig. 1. A $1 \times 1 \times 1$ cube design domain discretized by a $100 \times 100 \times 100$ FE mesh.

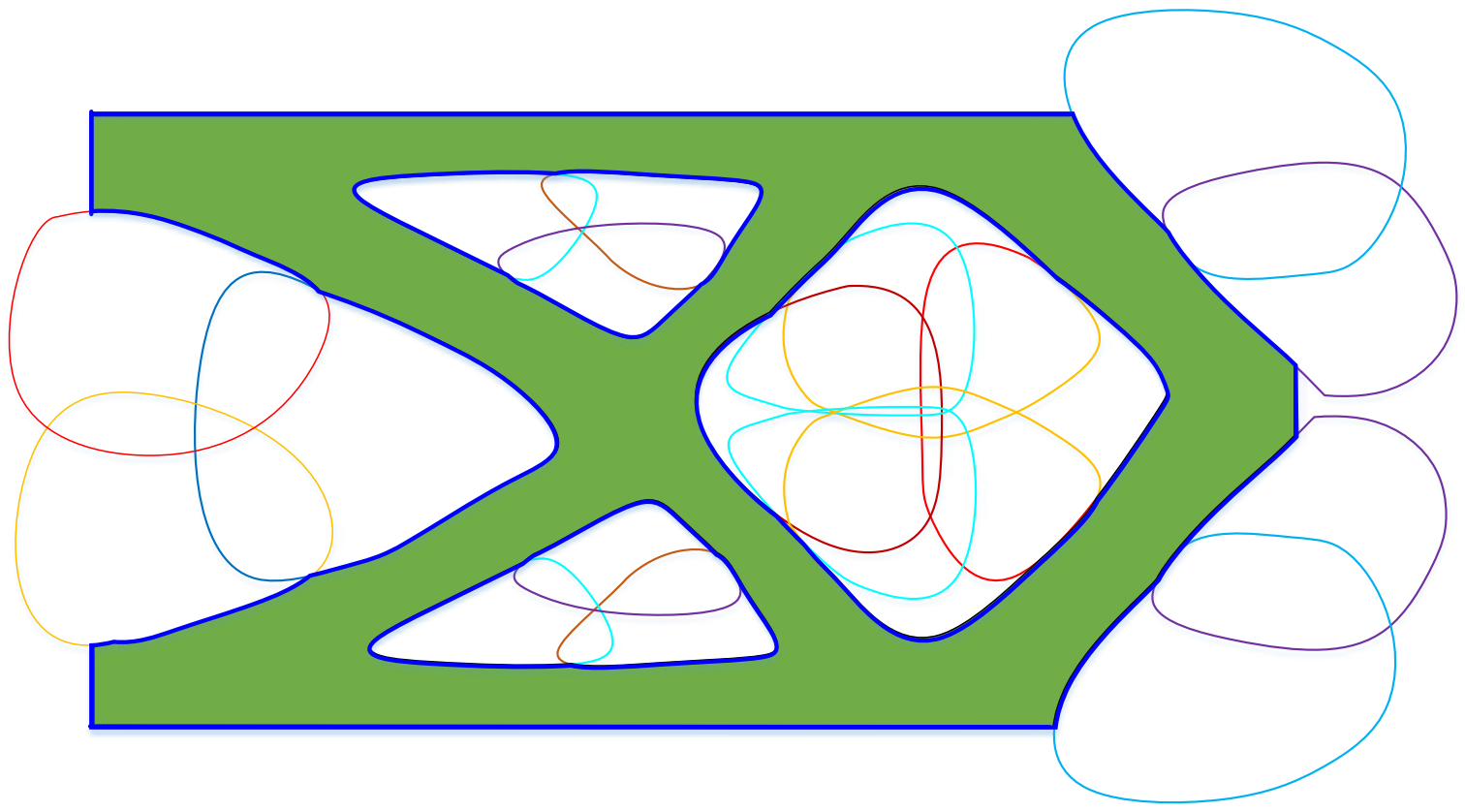

a. 2D case

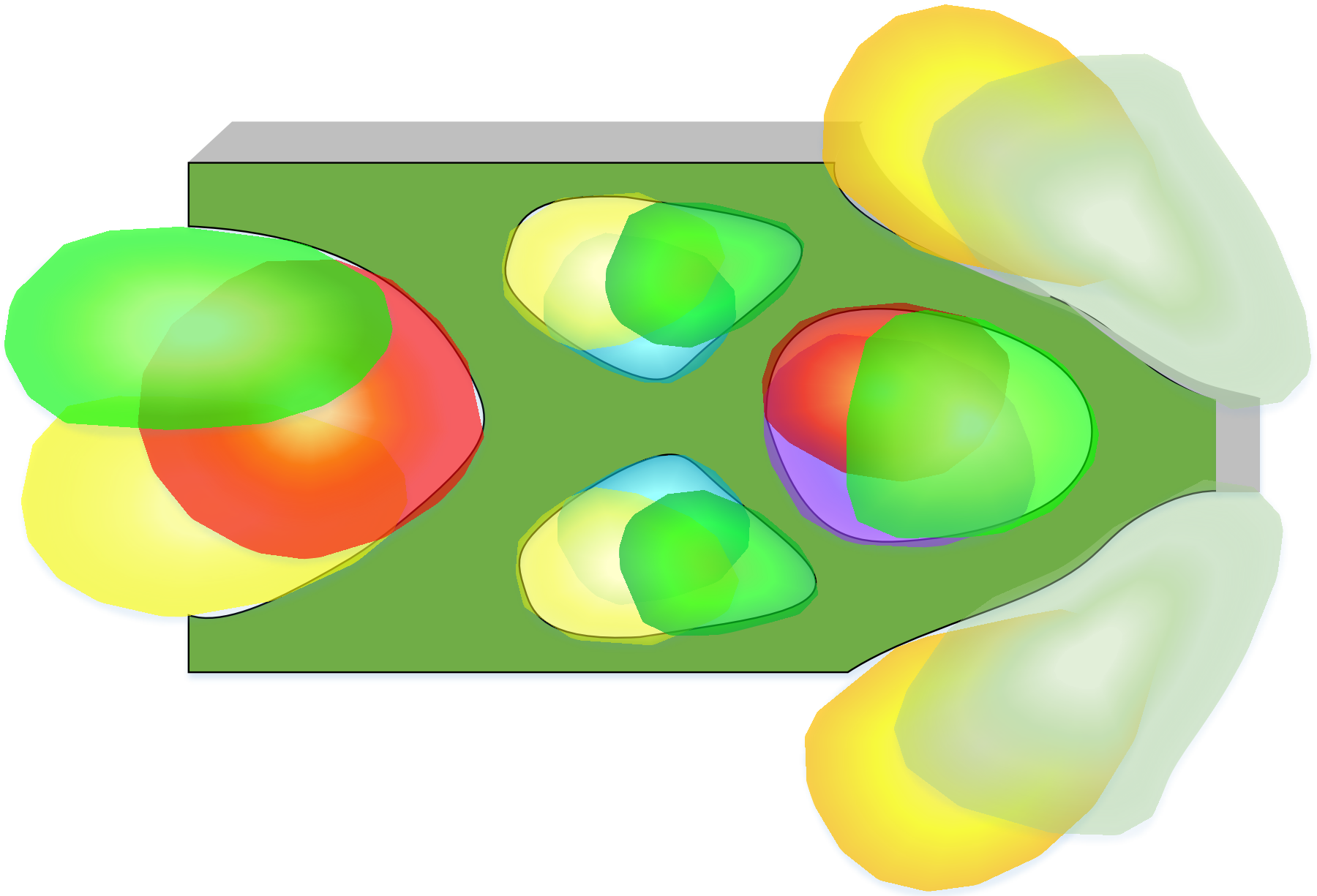

b. 3D case.

Fig. 2. The topology representation of a structure in the MMV approach.

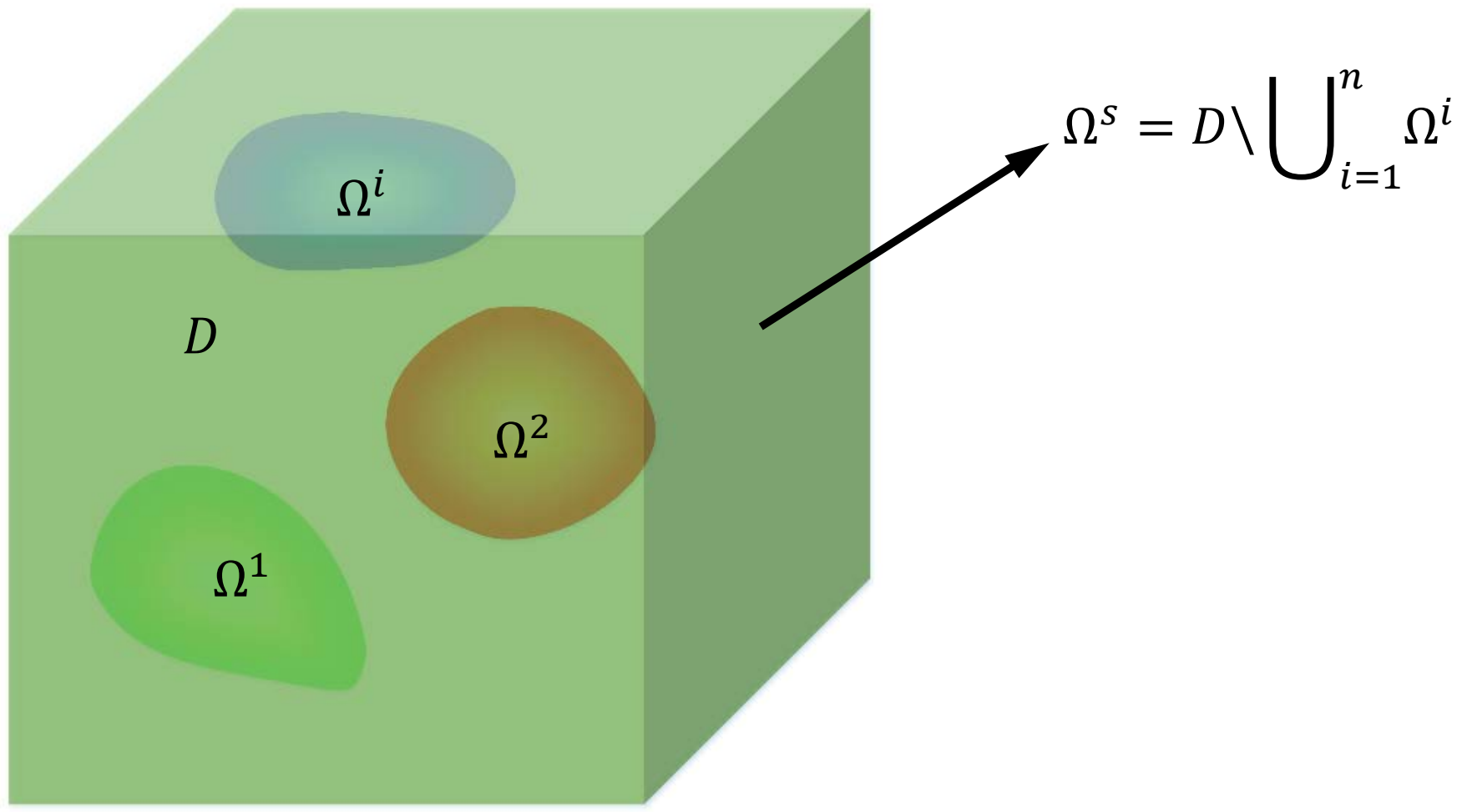

Fig. 3. The representation of $\Omega^s$ with use of a set of closed surfaces in the MMV approach.

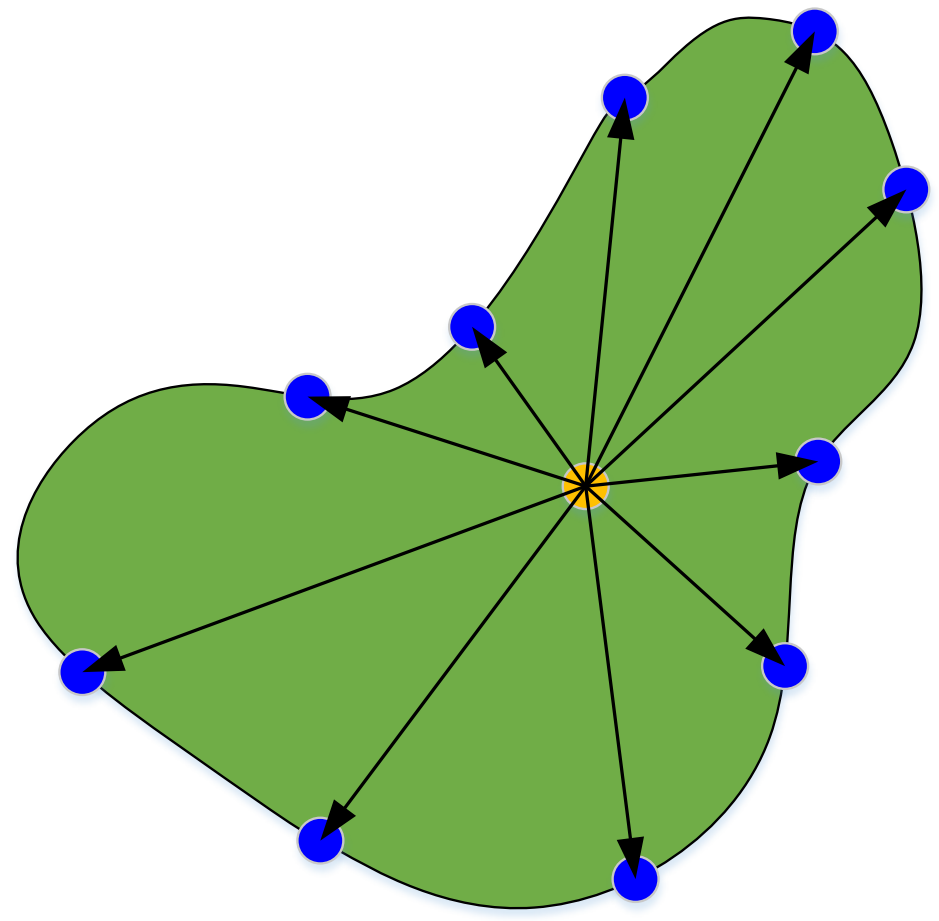

Fig. 4a. A star-shape region.

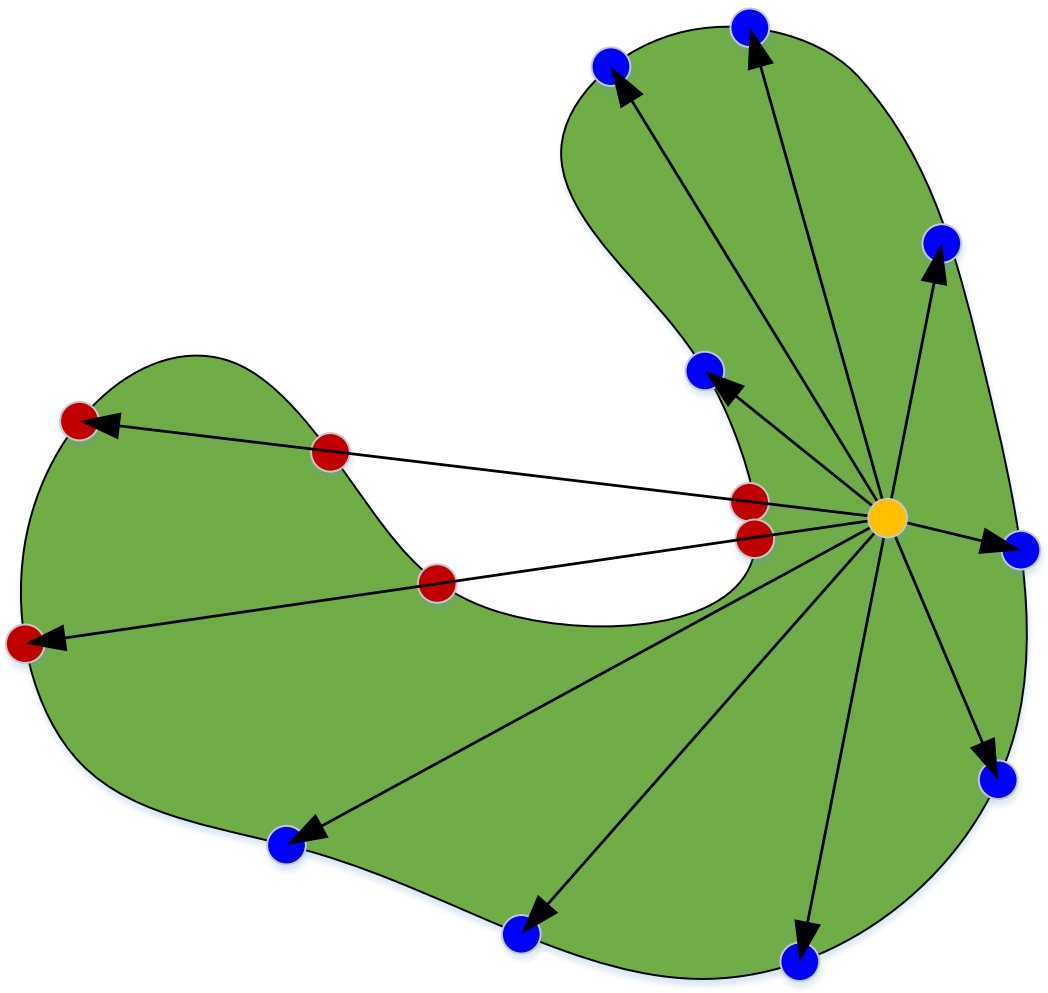

Fig. 4b. A non-star-shape region.

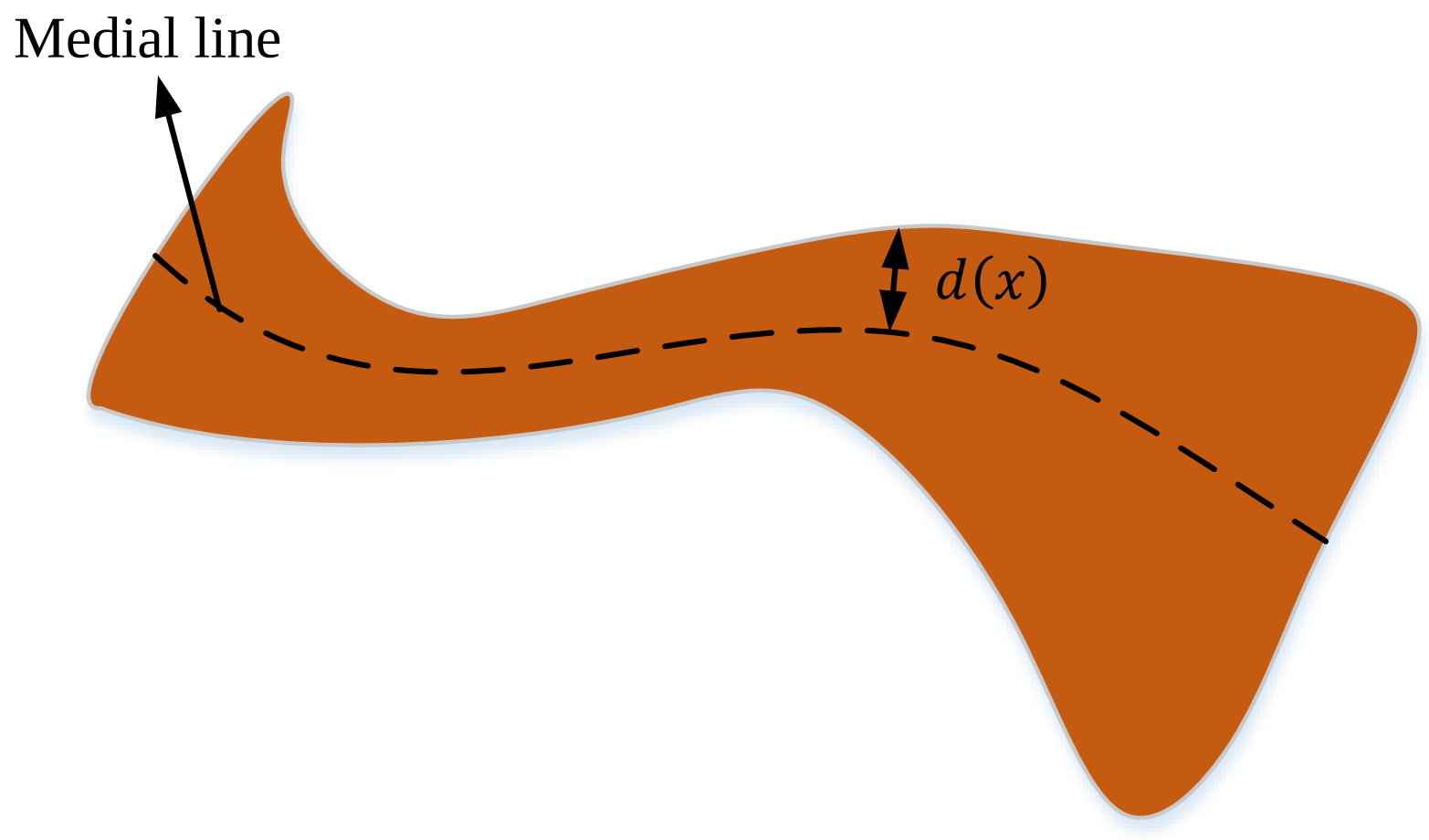

Fig. 5a. A 2D structure described by its medial line (skeleton) [24].

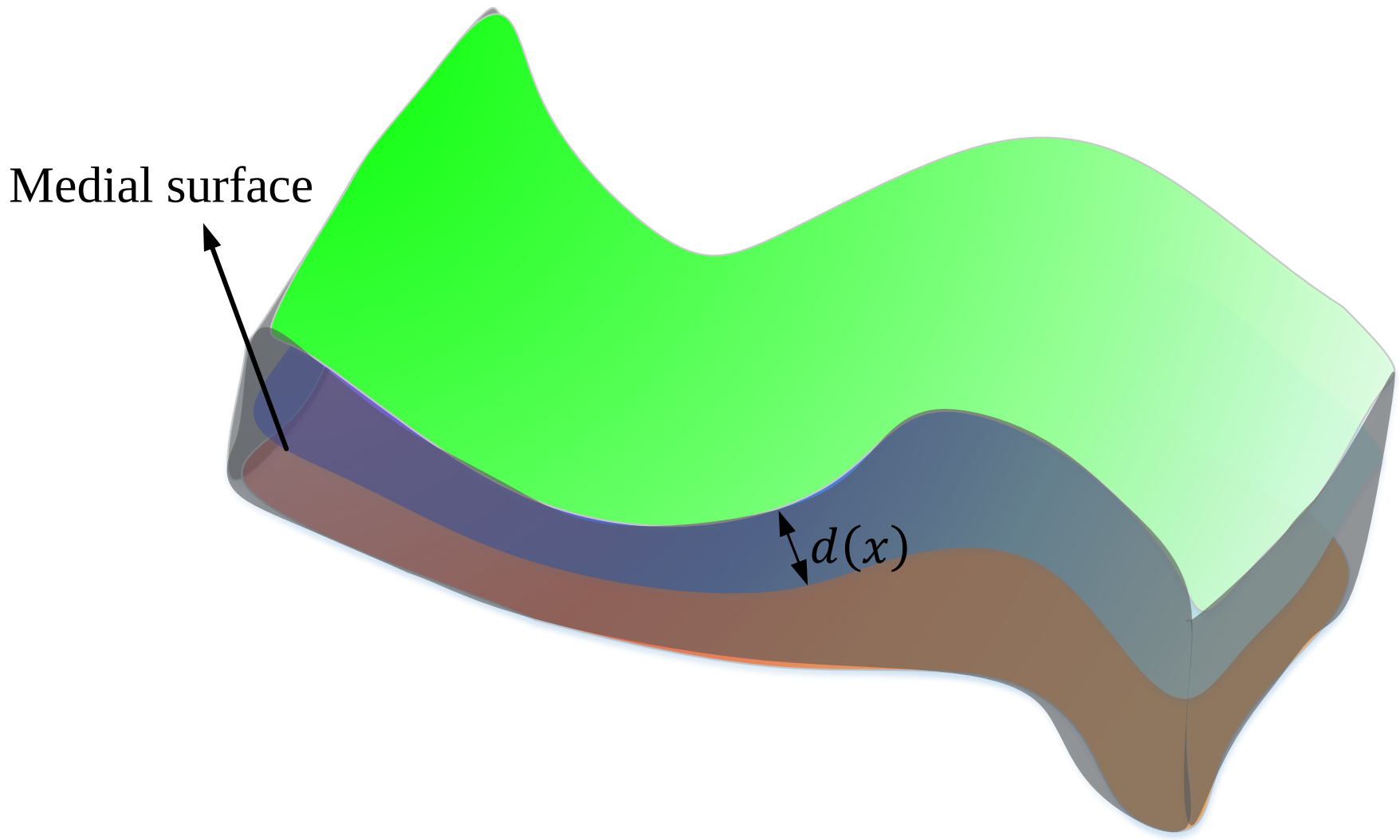

Fig. 5b. A 3D structure described by its medial surface (skeleton) [24].

Fig. 5. Geometry description of a component with its curved medial line/surface.

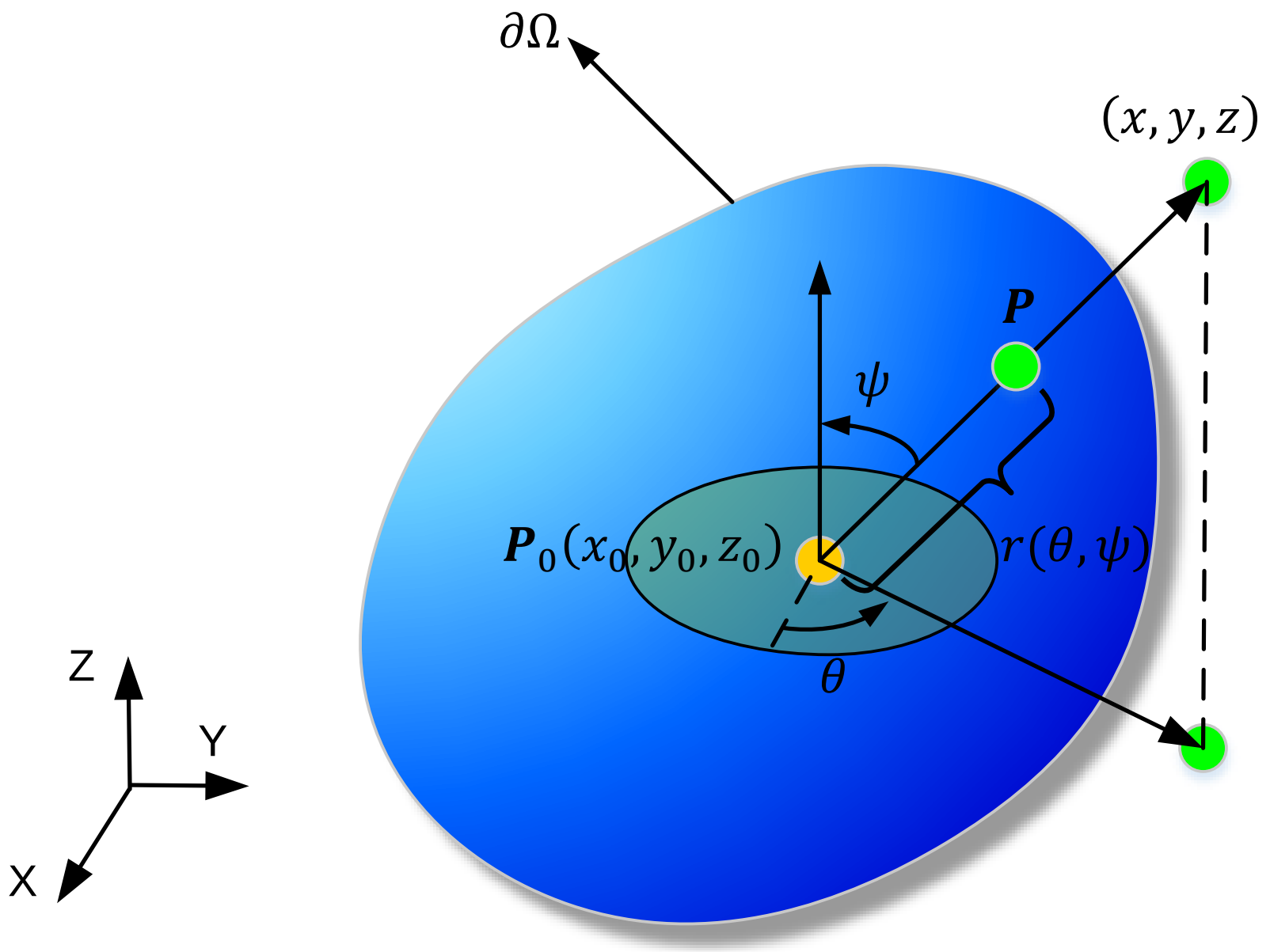

Fig. 6. The geometry representation of a 3D star-shaped region with use of Eq (3.1).

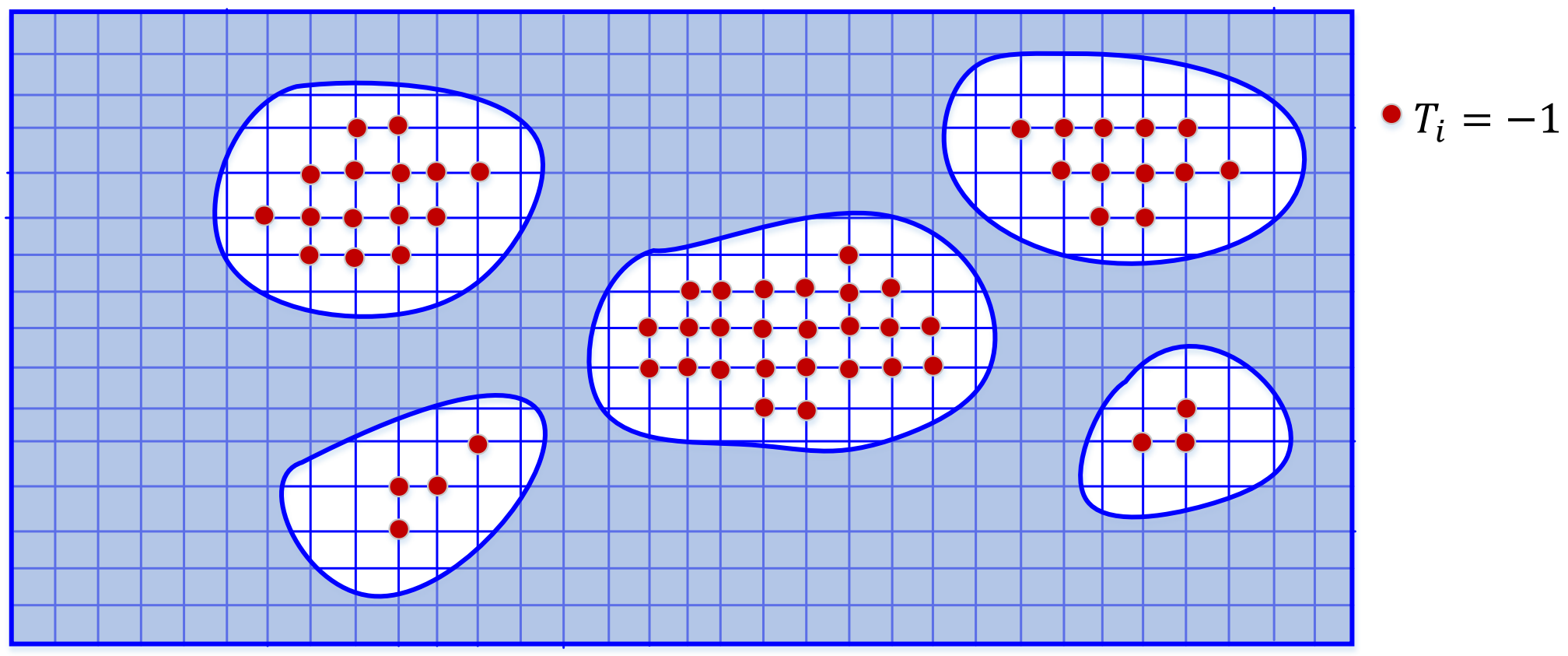

Fig. 7. A schematic illustration of the DOF removal technique in FE analysis (MMV approach).

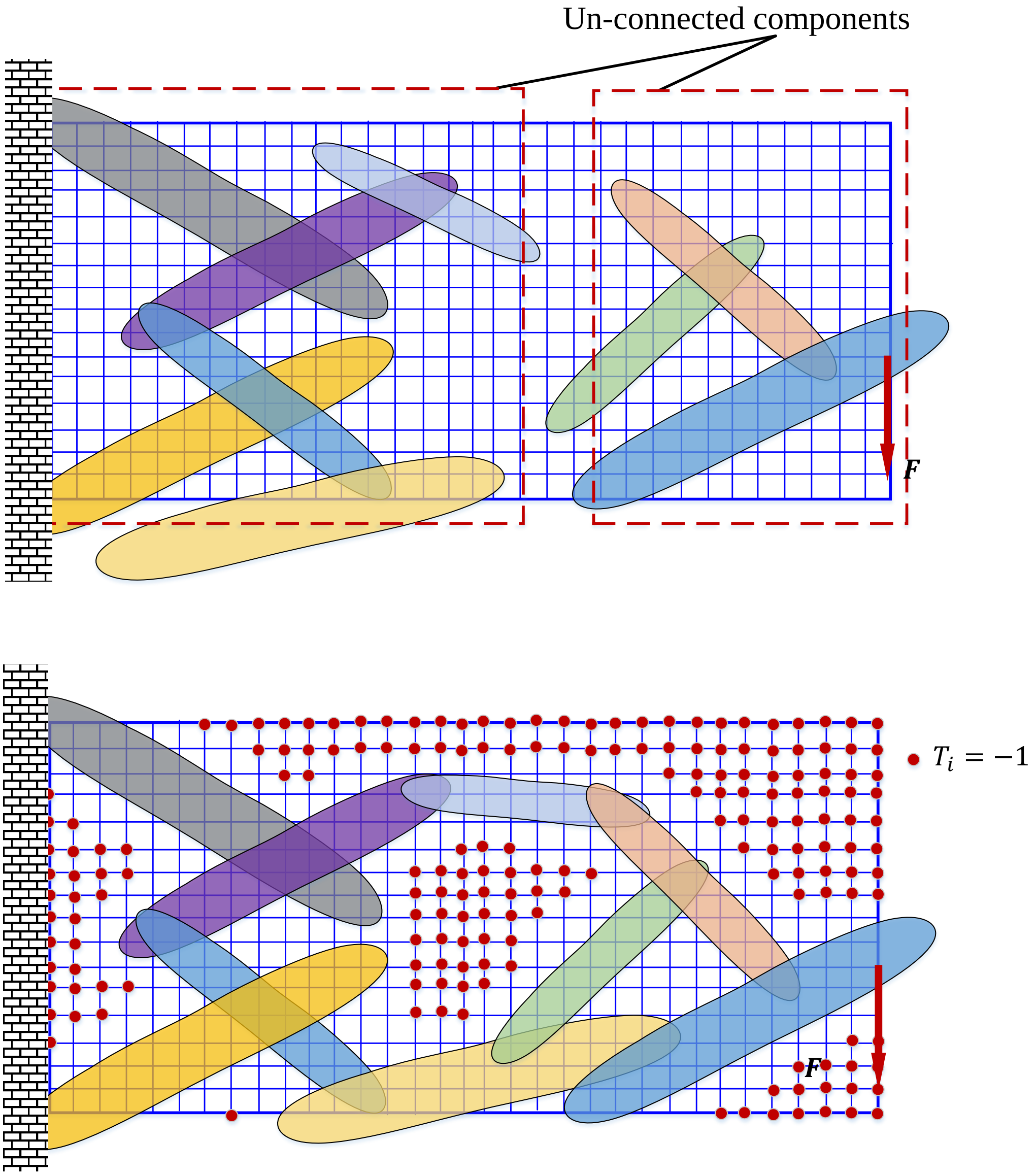

Fig. 8. A schematic illustration of the DOF removal technique in FE analysis

(MMC approach).

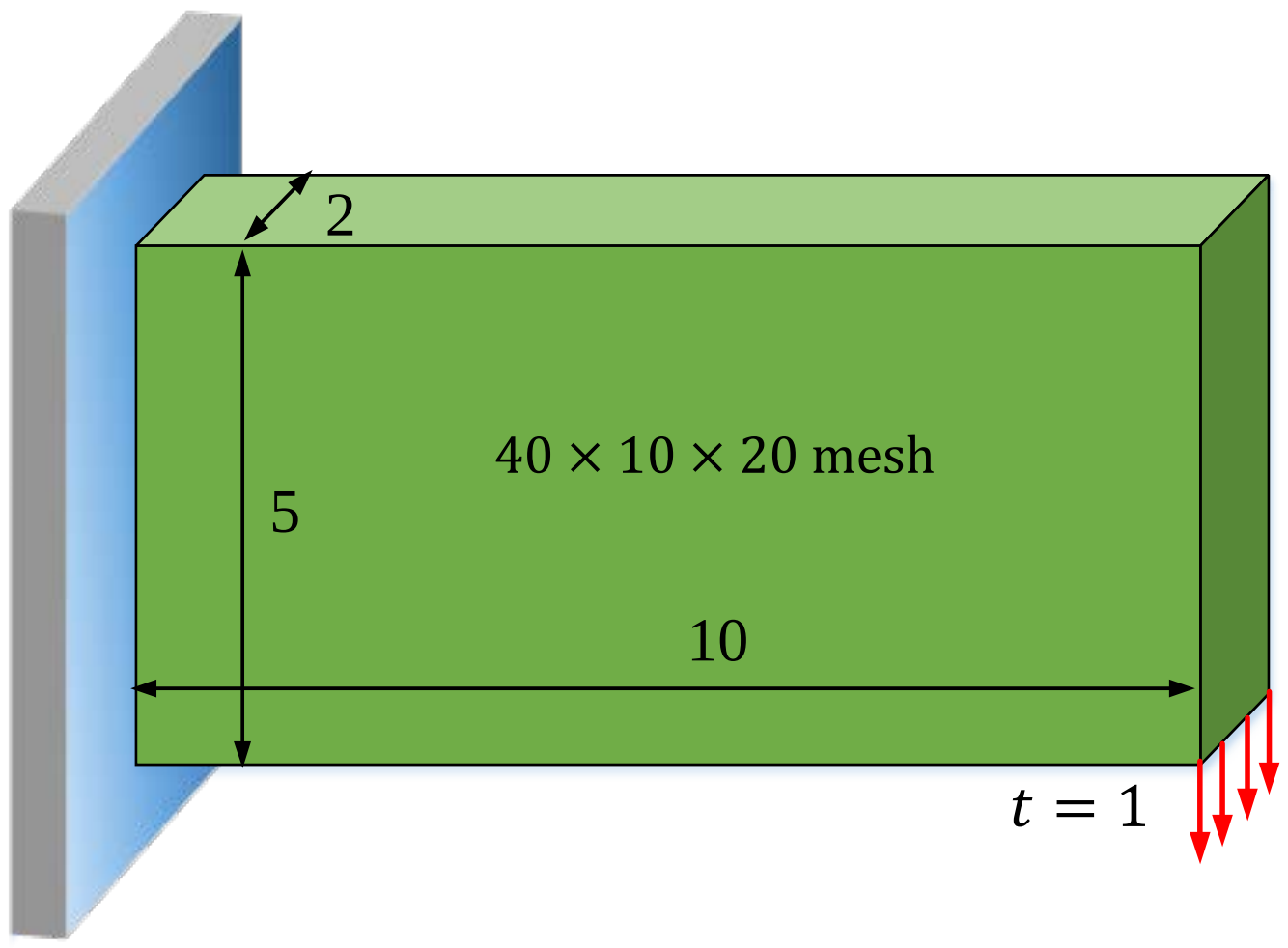

Fig. 9 The short cantilever beam example.

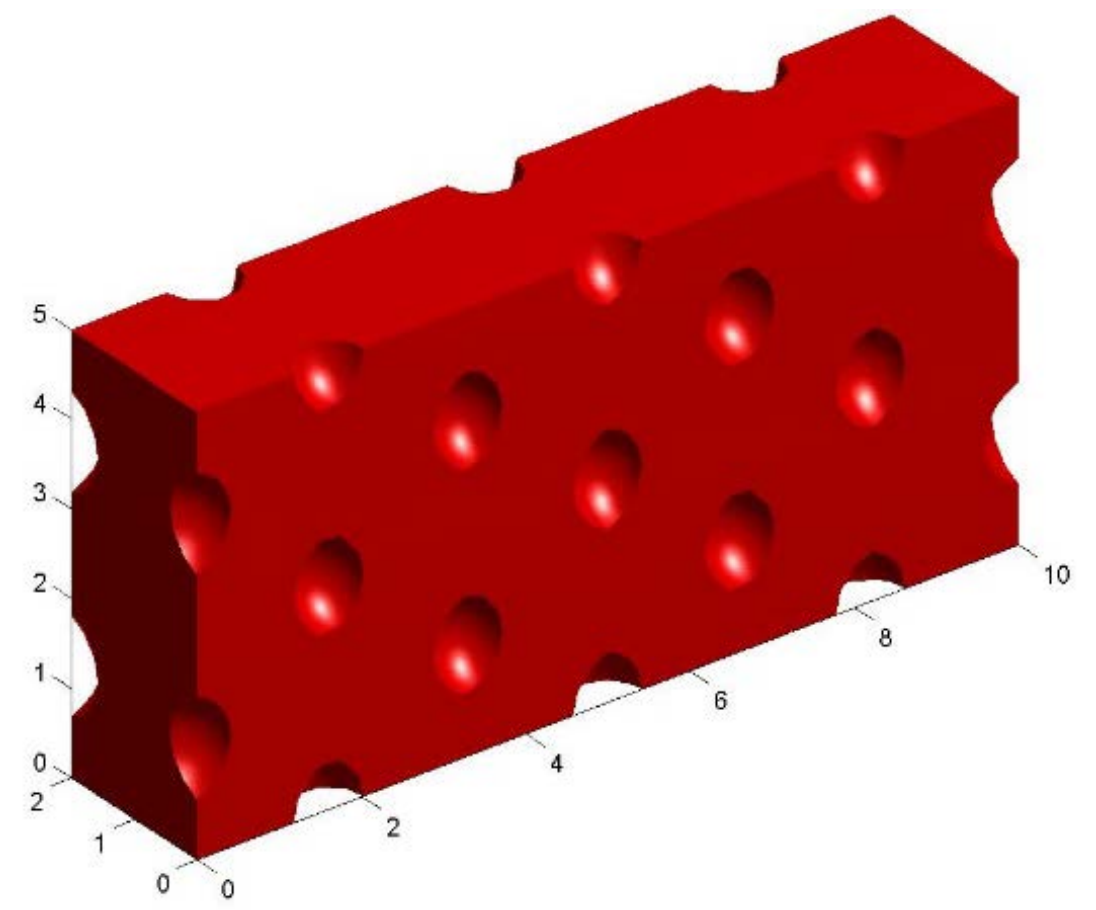

a. Voids described by a Hermite interpolation.

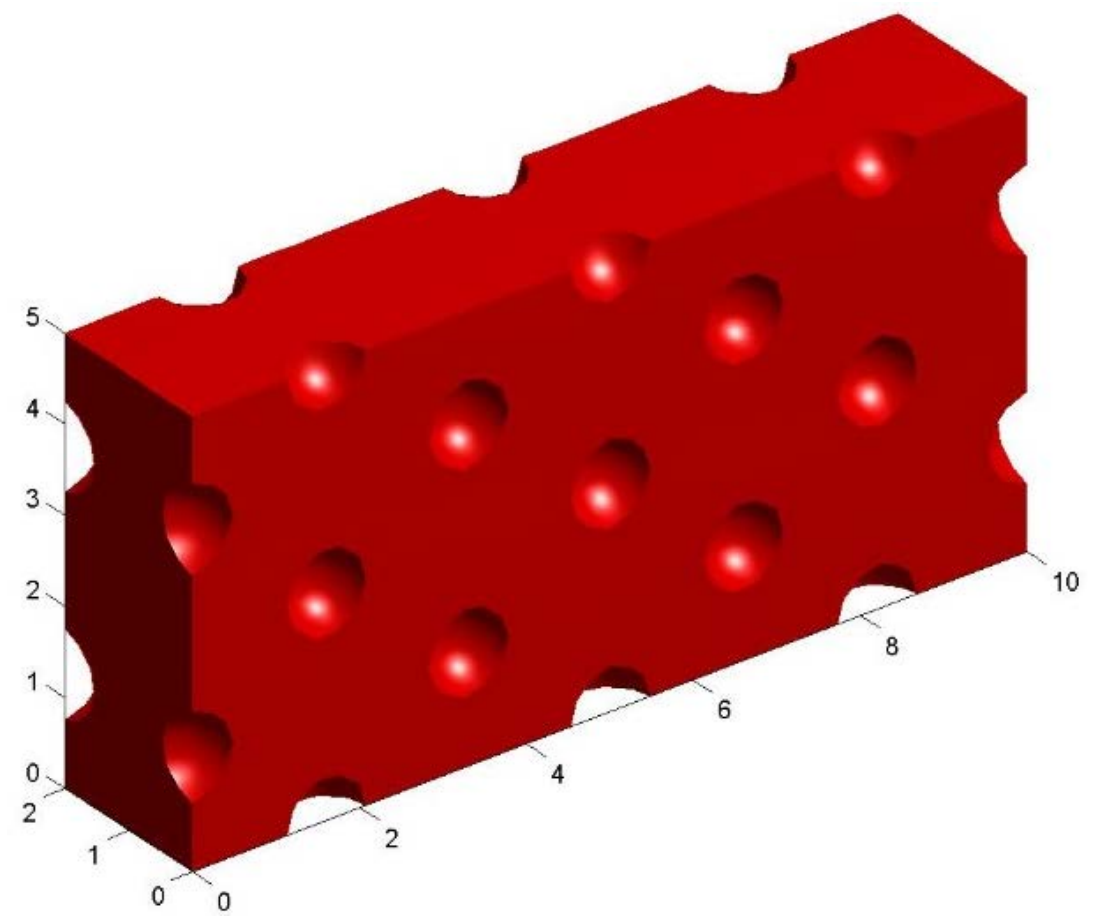

b. Voids described by a NURBS interpolation.

Fig. 10 The initial designs for the short cantilever beam example.

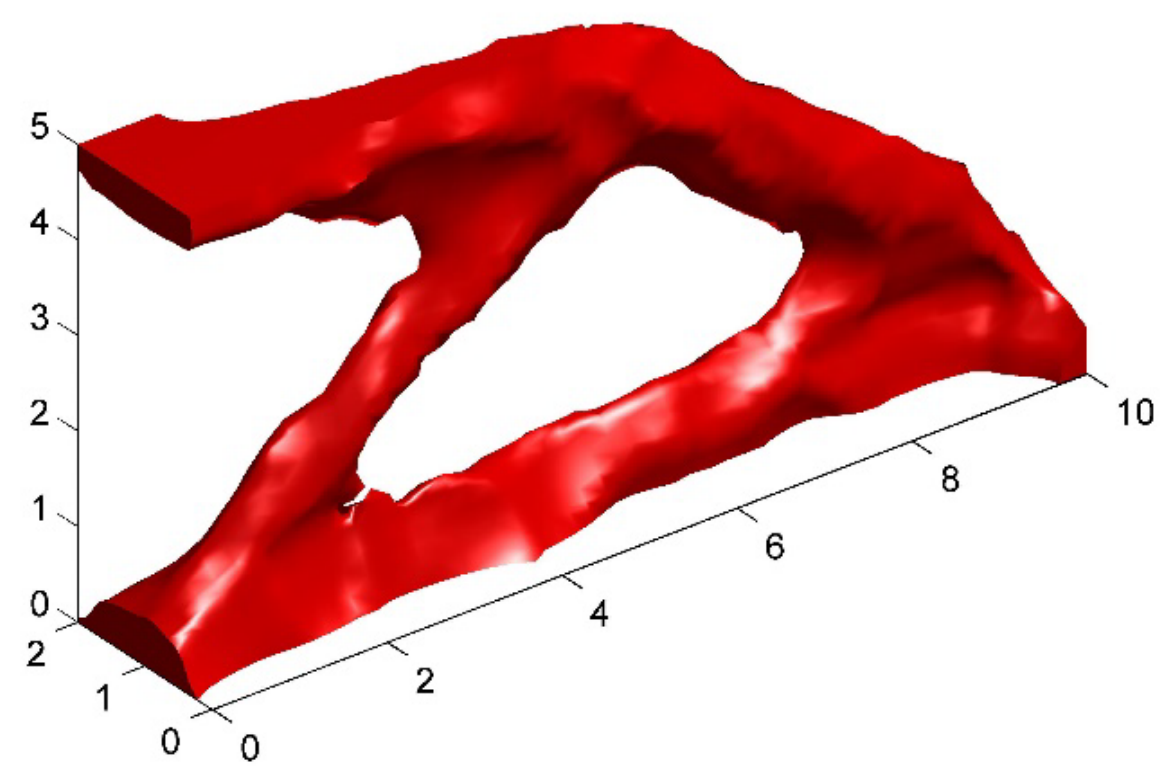

a. Voids described by Hermite interpolation.

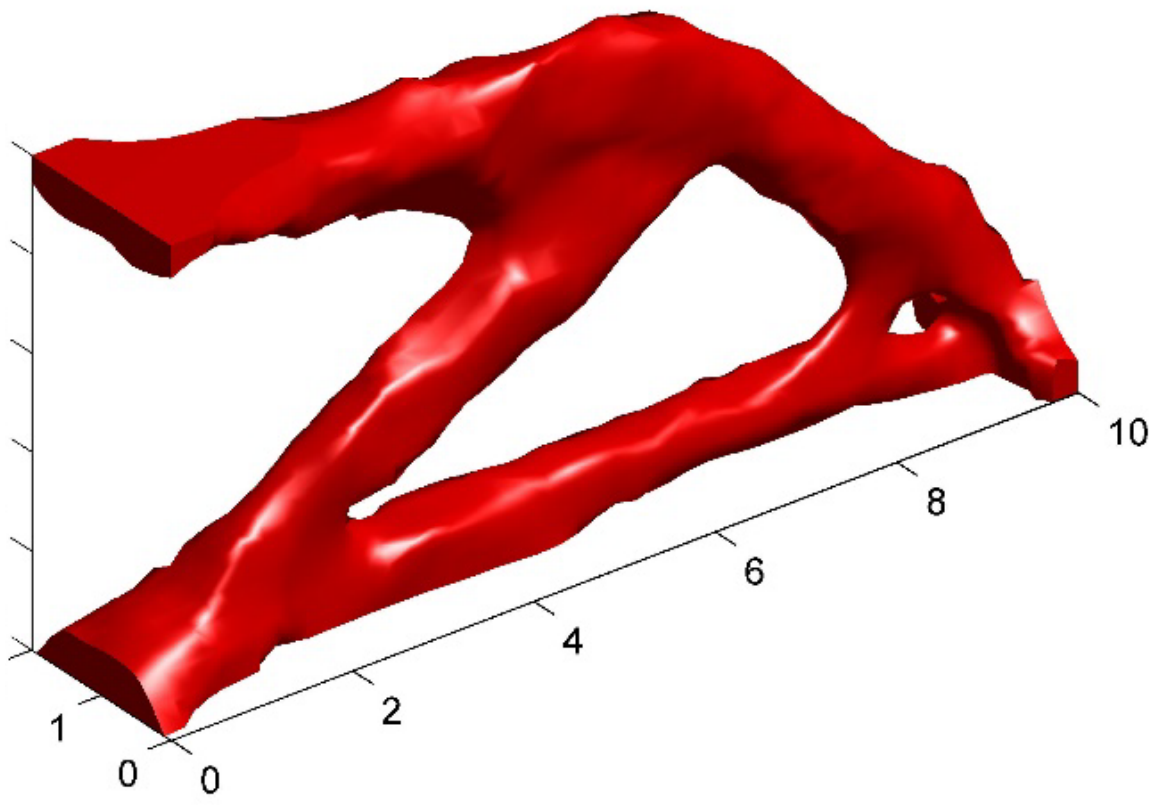

b. Voids described by NURBS interpolation.

Fig. 11 Optimized structures for the short cantilever beam example.

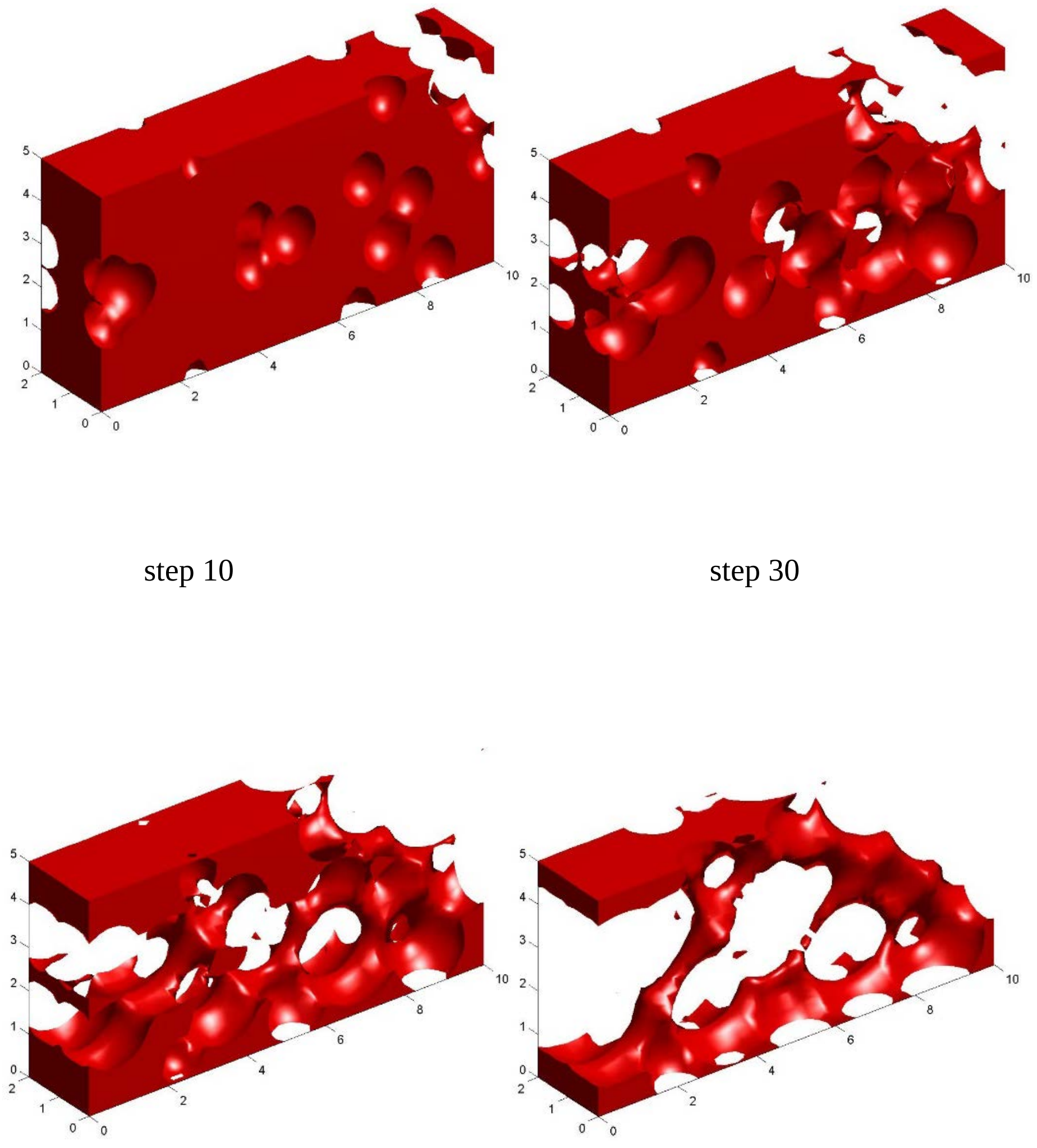

step 10 step 30

step 50 step 100

a. Voids described by Hermite interpolation.

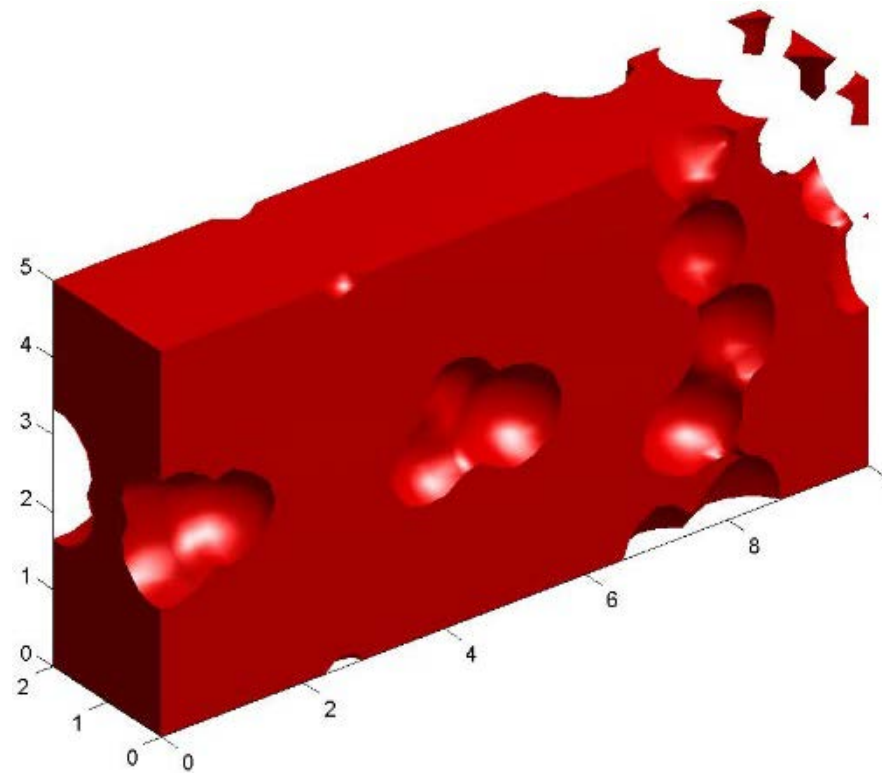

step 10

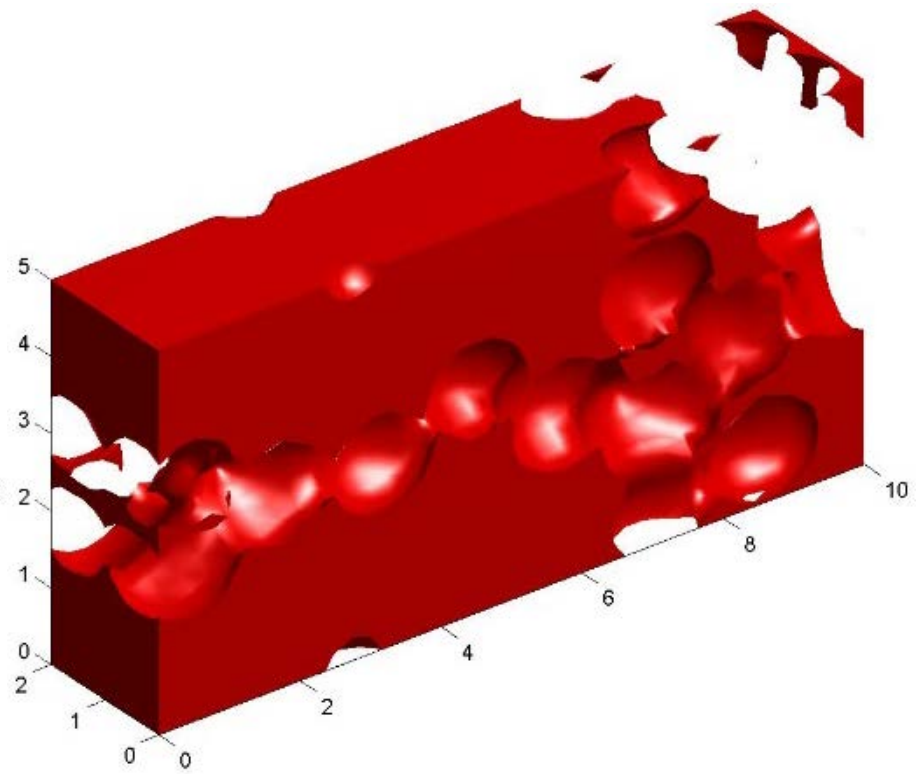

step 30

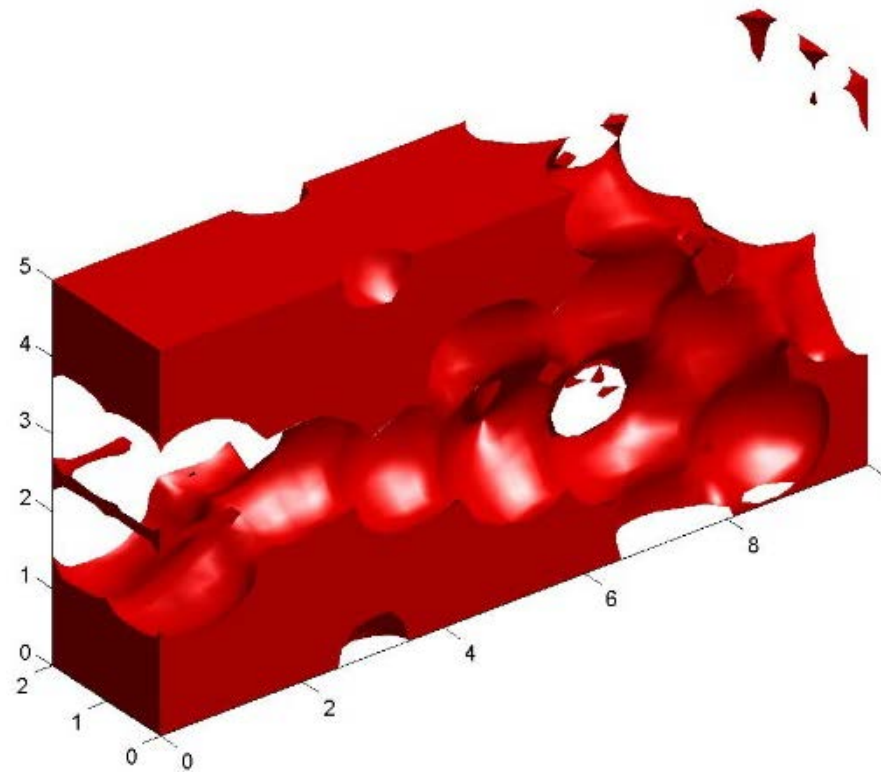

step 50

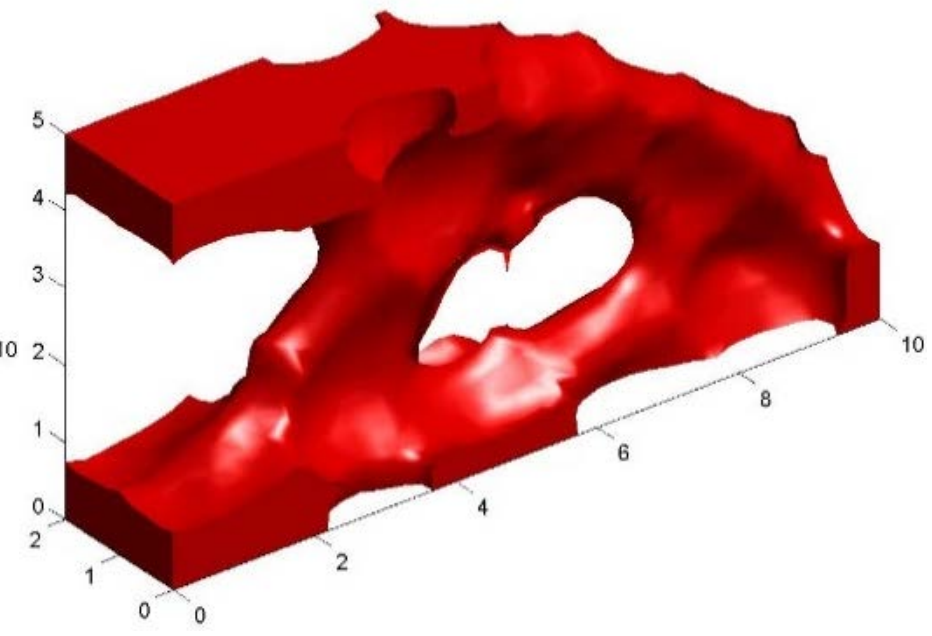

step 100

b. Voids described by NURBS interpolation.

Fig. 12 Some intermediate iteration steps for the short cantilever beam example.

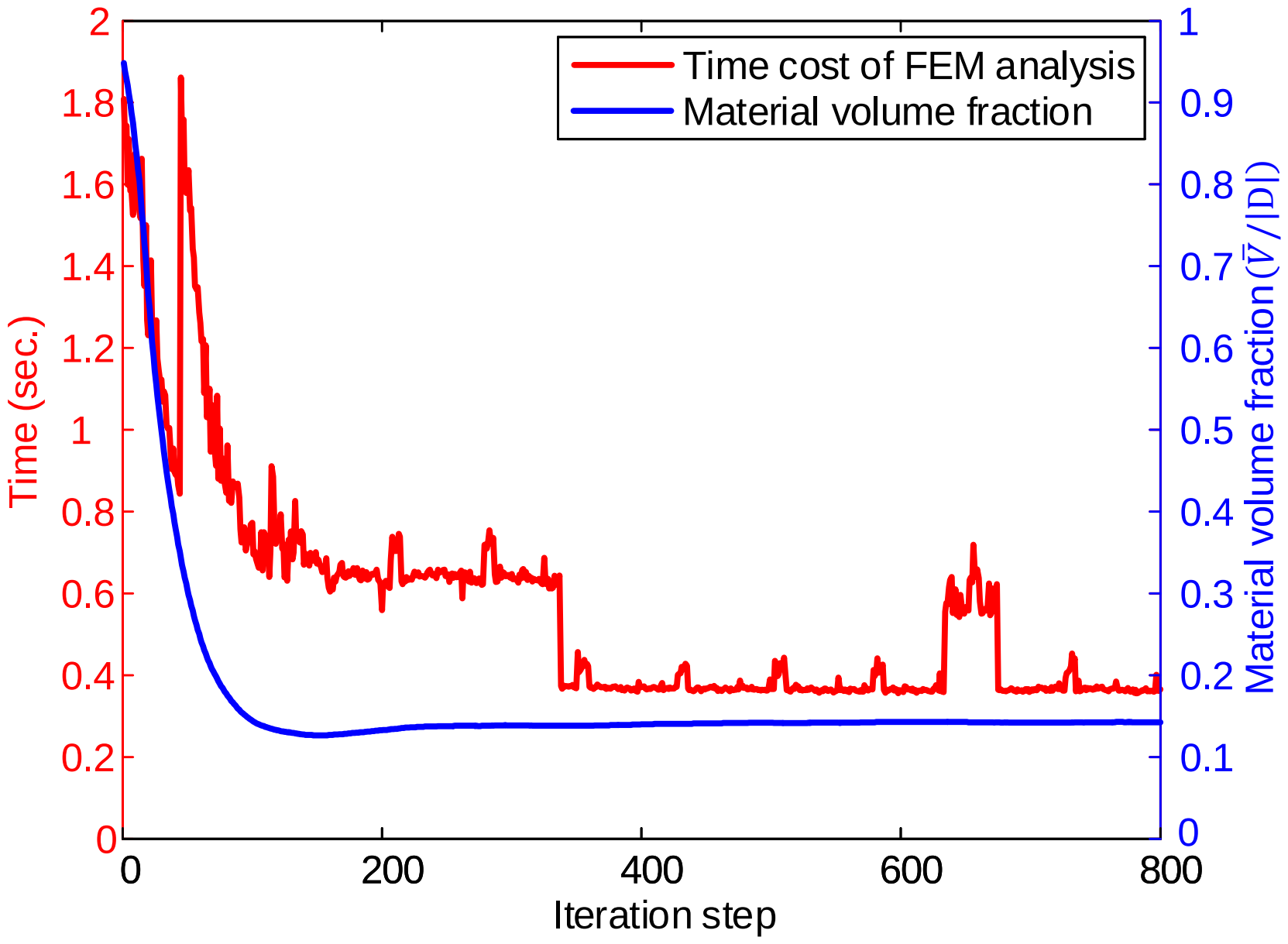

a. Voids described by Hermite interpolation.

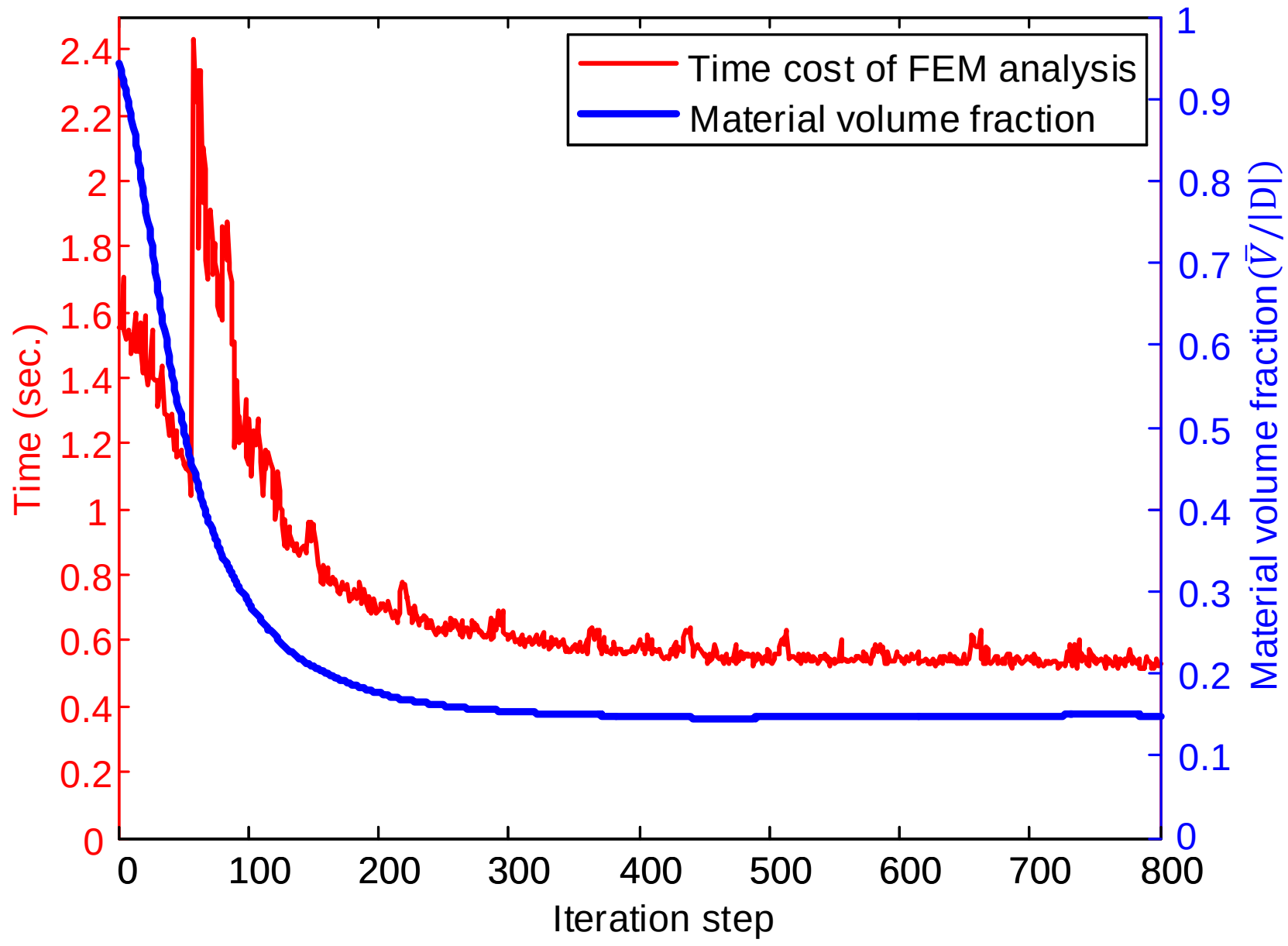

b. Voids described by NURBS interpolation.

Fig. 13 The CPU time for the FE analysis of the short cantilever beam example.

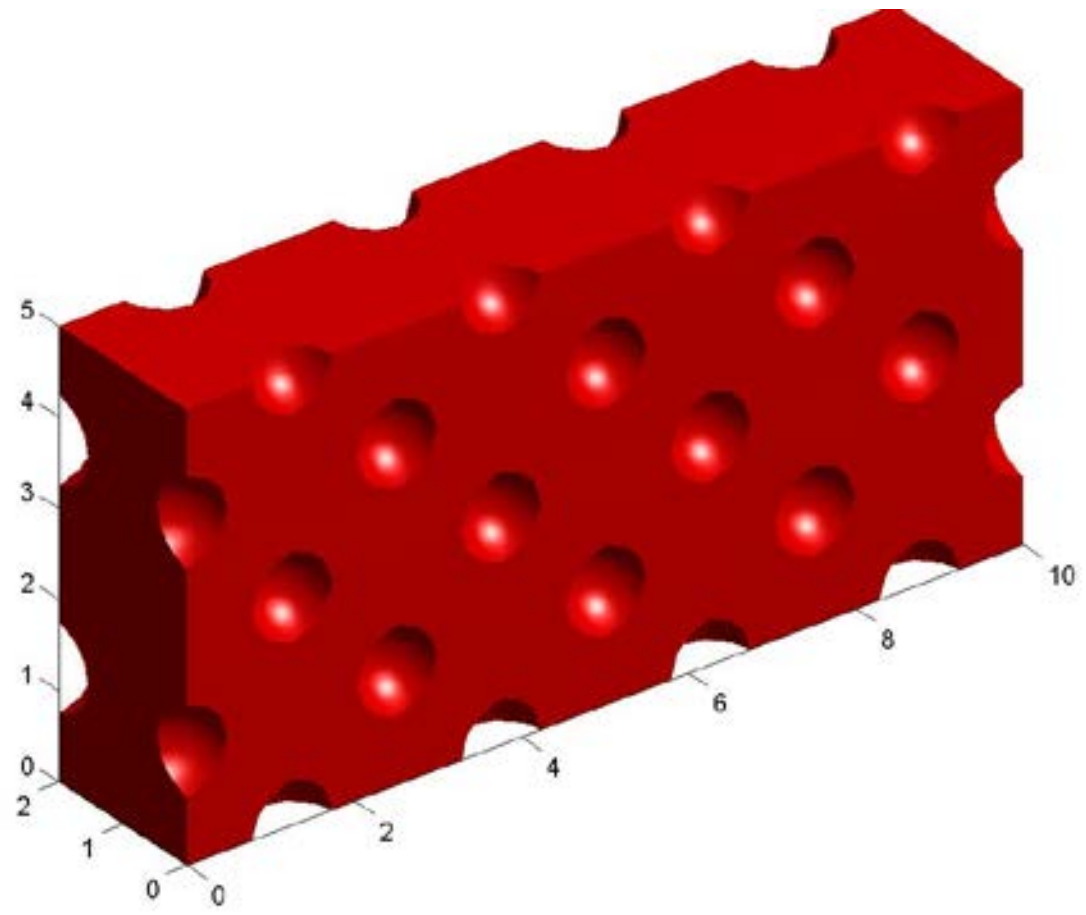

Fig. 14 The initial design for the short cantilever beam example (with 55 voids).

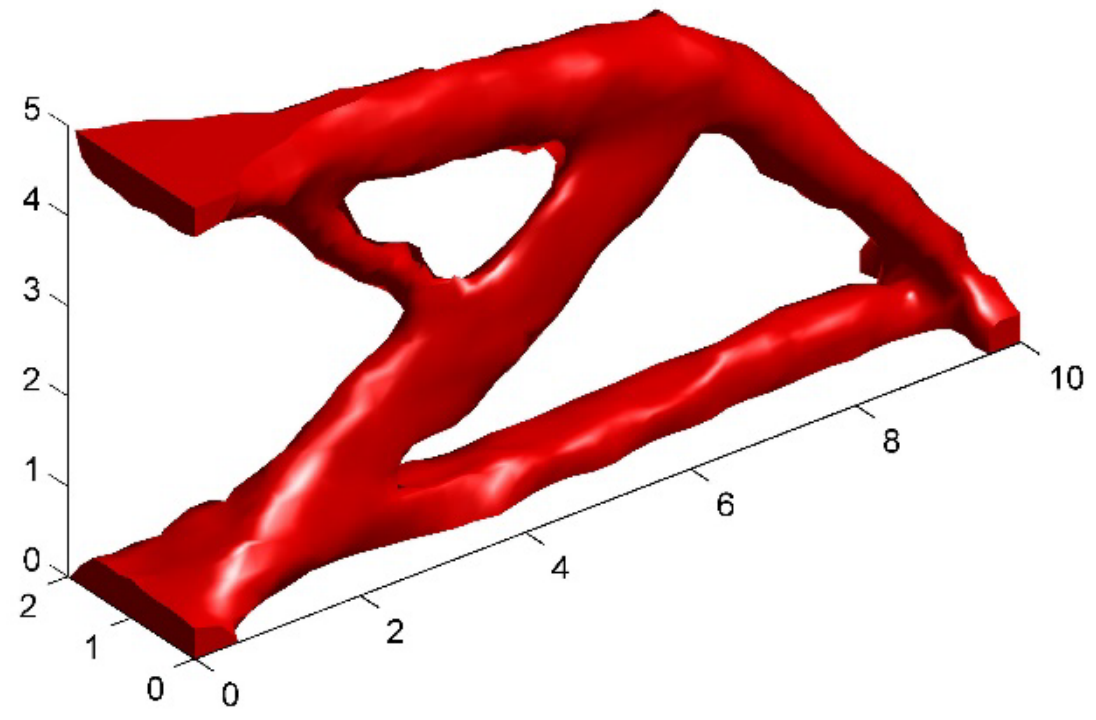

Fig. 15 Optimized structures for the short cantilever beam example.
(with the initial design shown in Fig. 14)

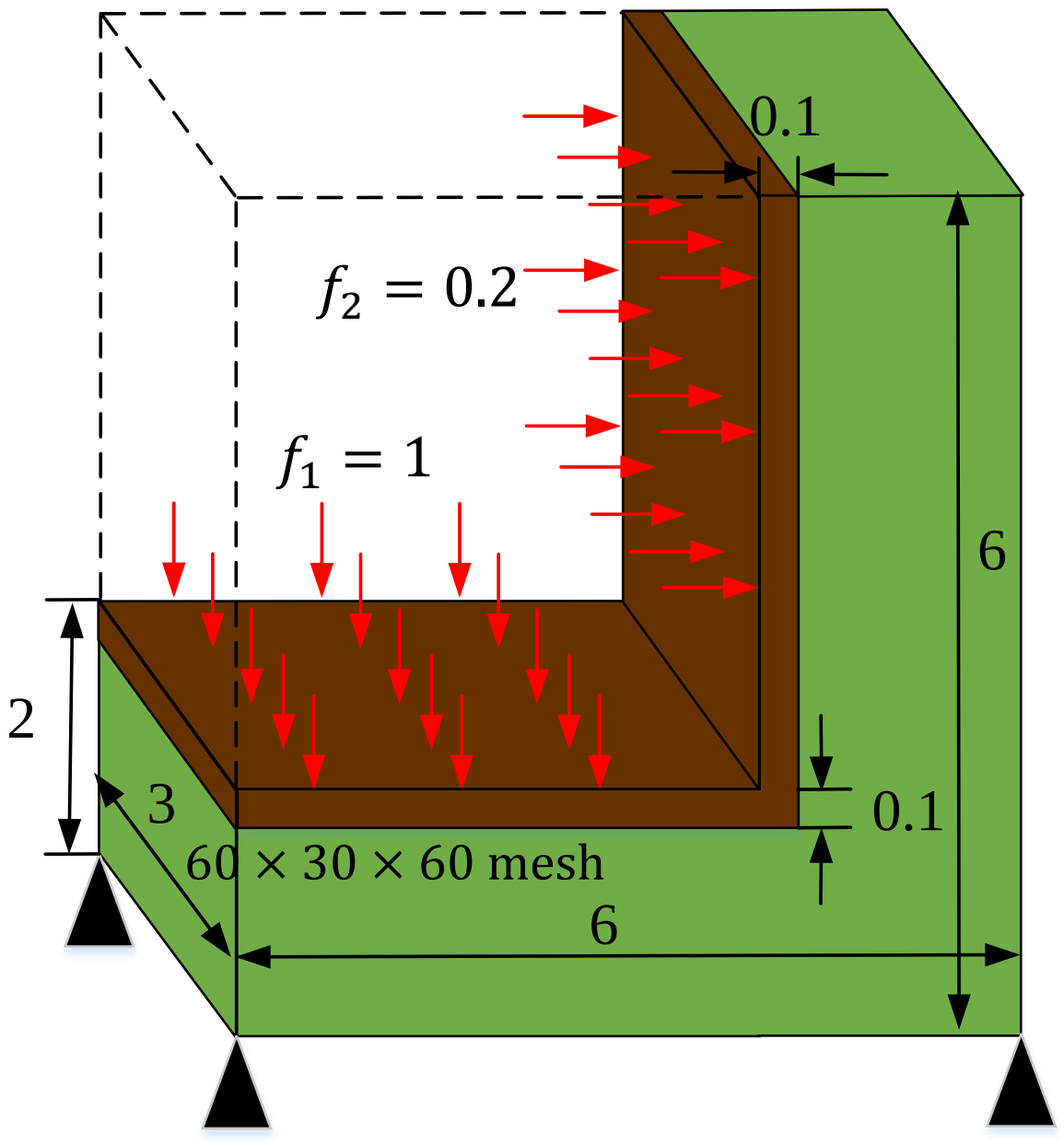

Fig. 16 The L-shape chair example.

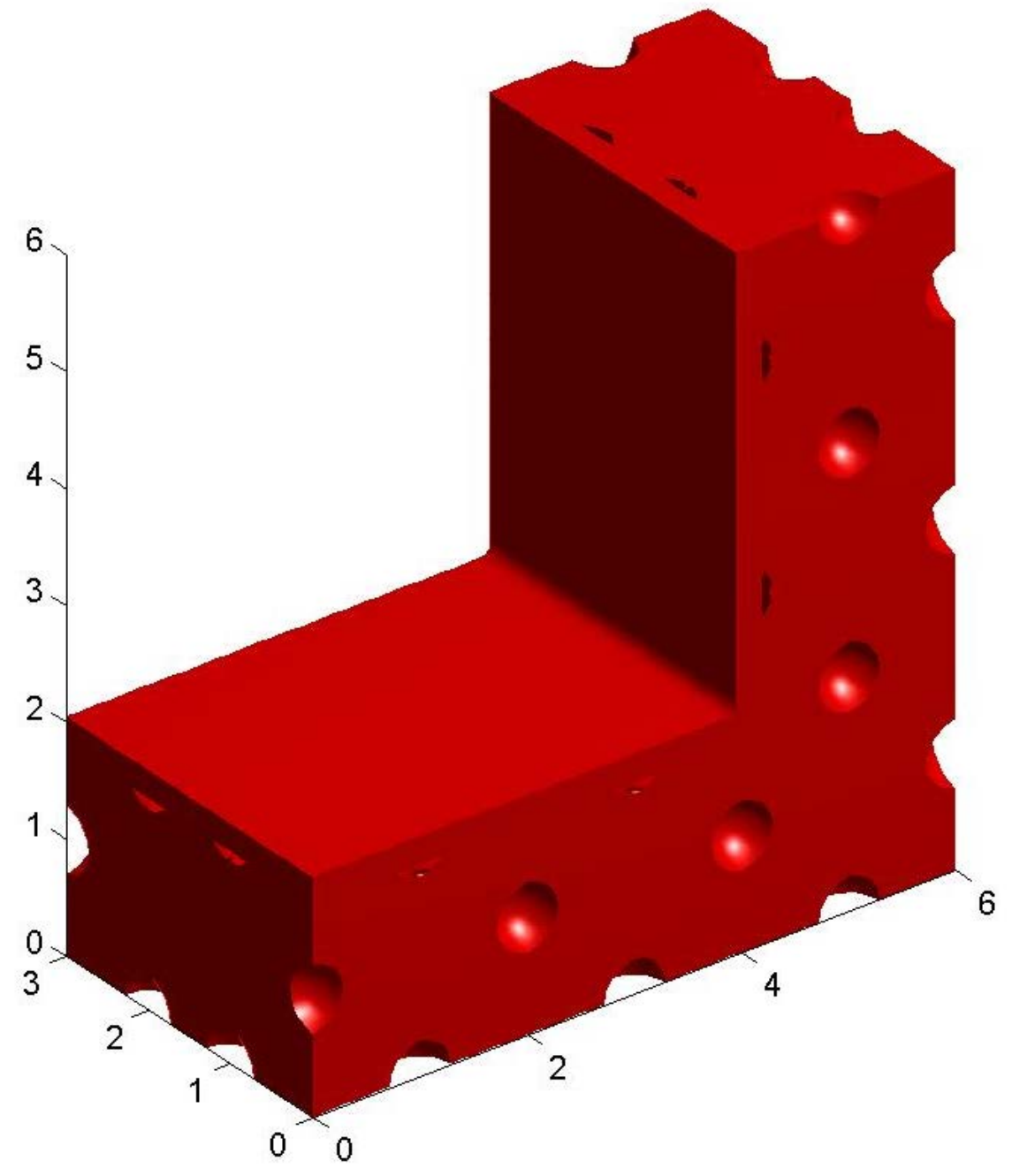

Fig. 17 The initial design for the L-shape chair example.

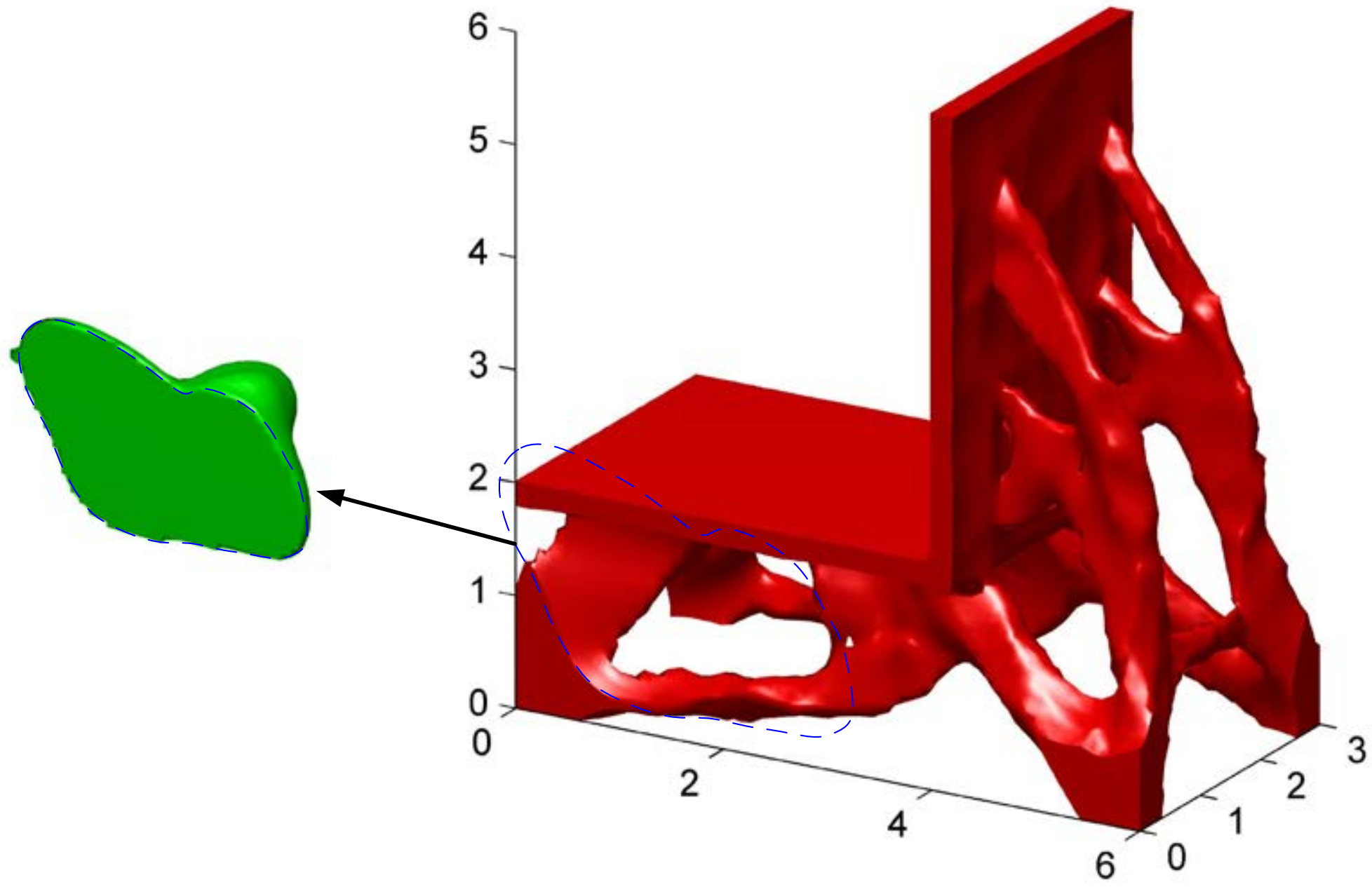

Fig. 18 The optimized structure of the L-shape chair example.

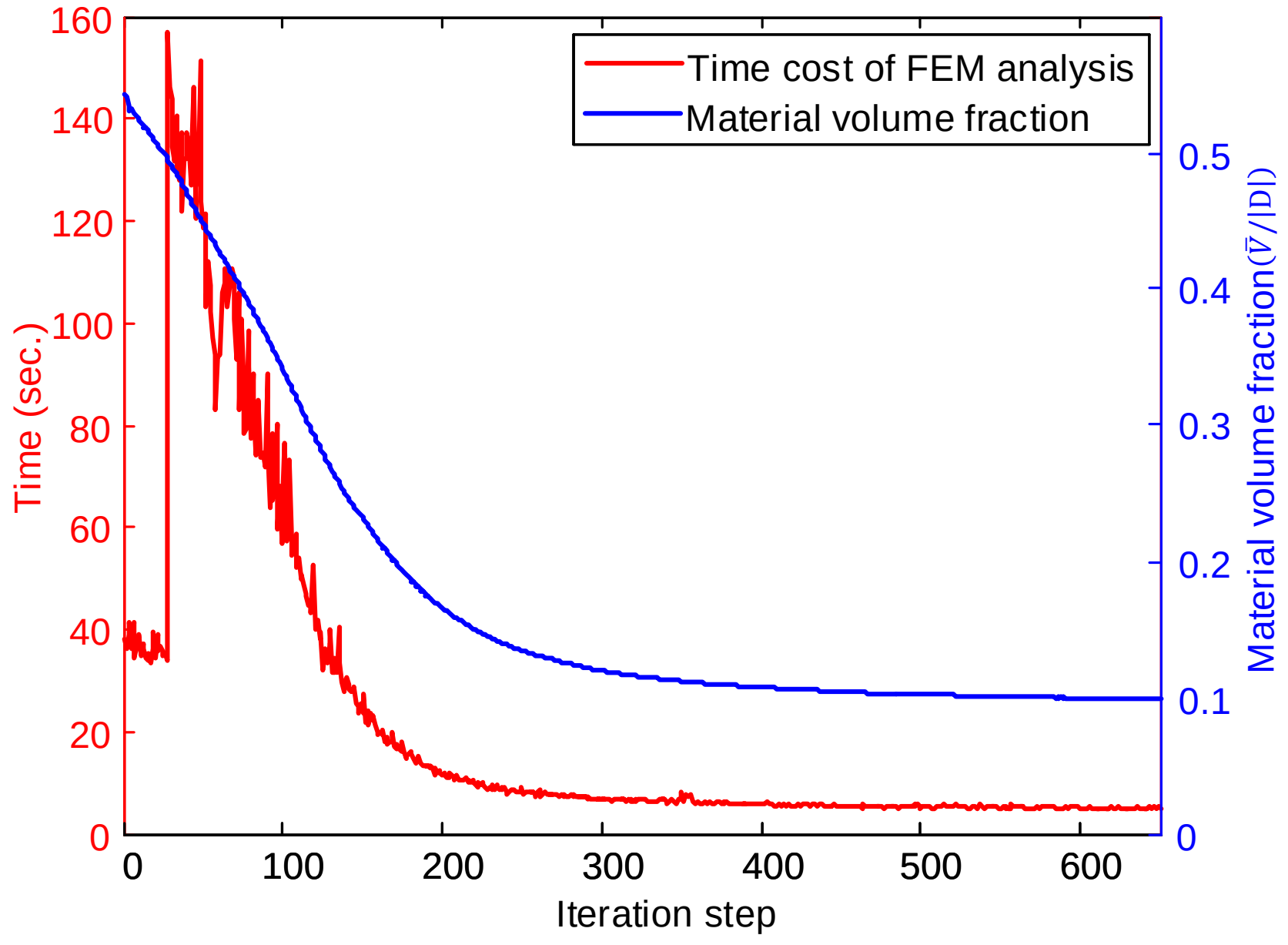

Fig. 19 The CPU time for the FE analysis of the L-shape chair example.

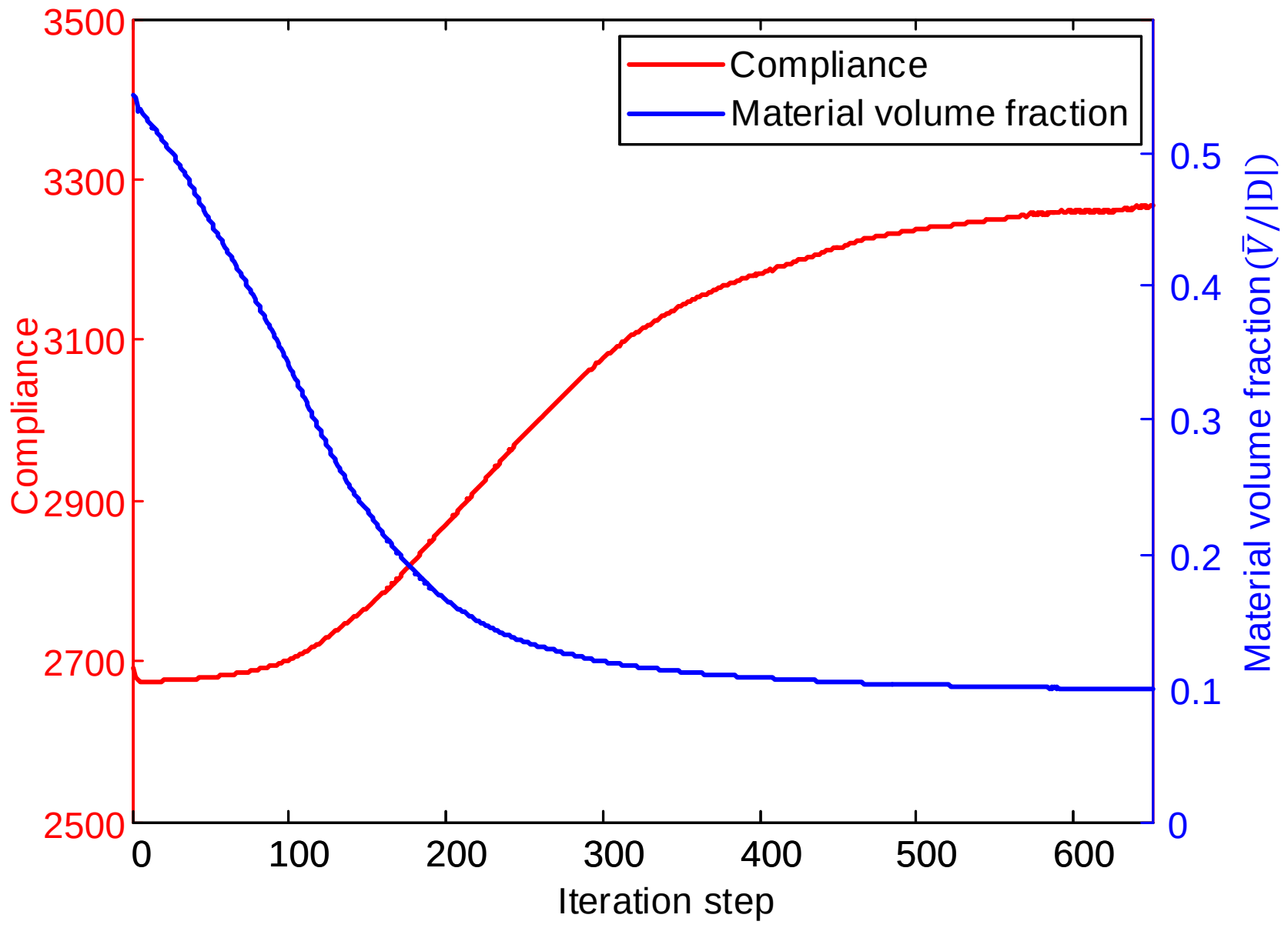

Fig. 20 Convergence history of the L-shape chair example.

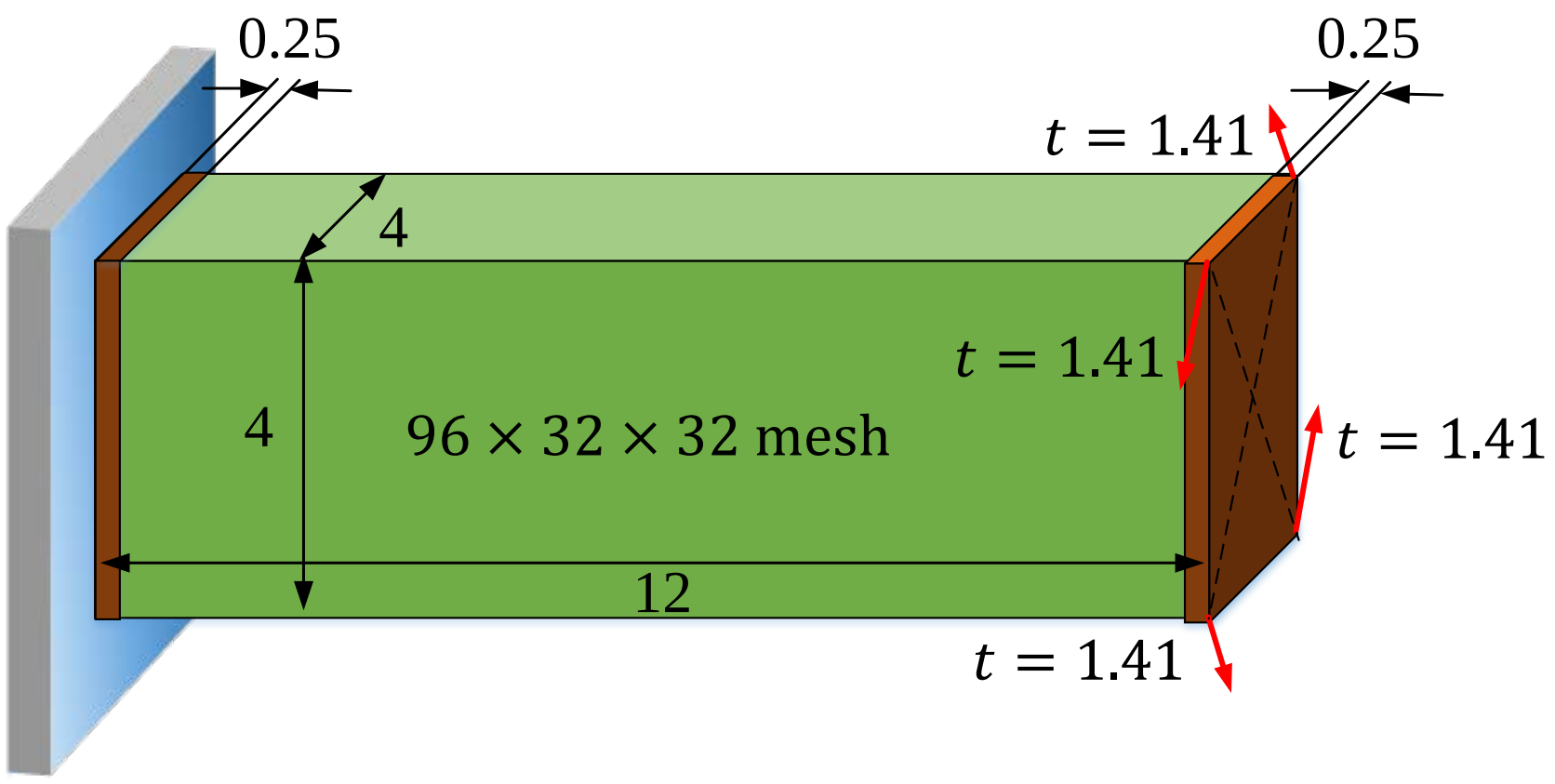

Fig. 21 The torsion beam example.

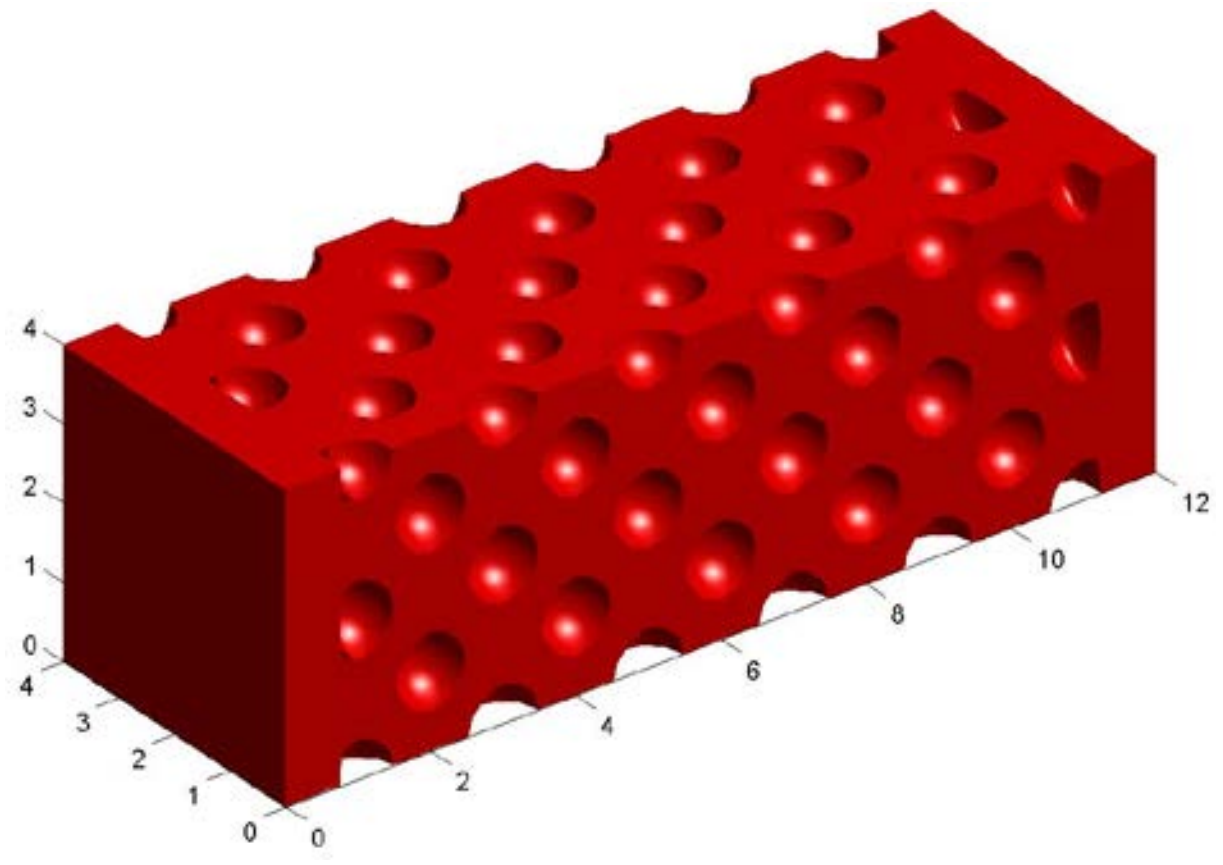

Fig. 22a The initial design for the torsion beam example by the MMV approach.

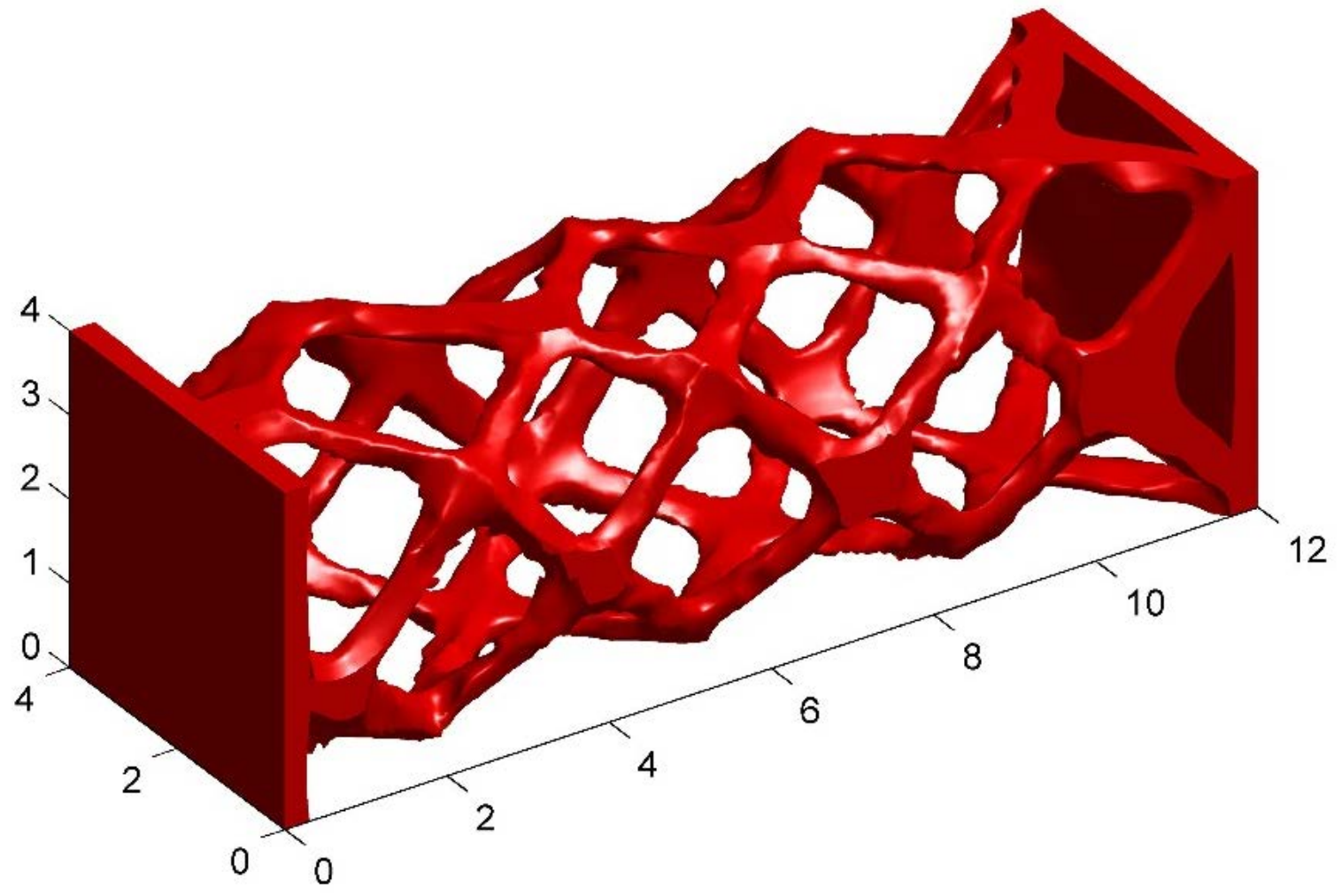

Fig. 22b Optimized structures for the torsion beam example obtained by the MMV approach.

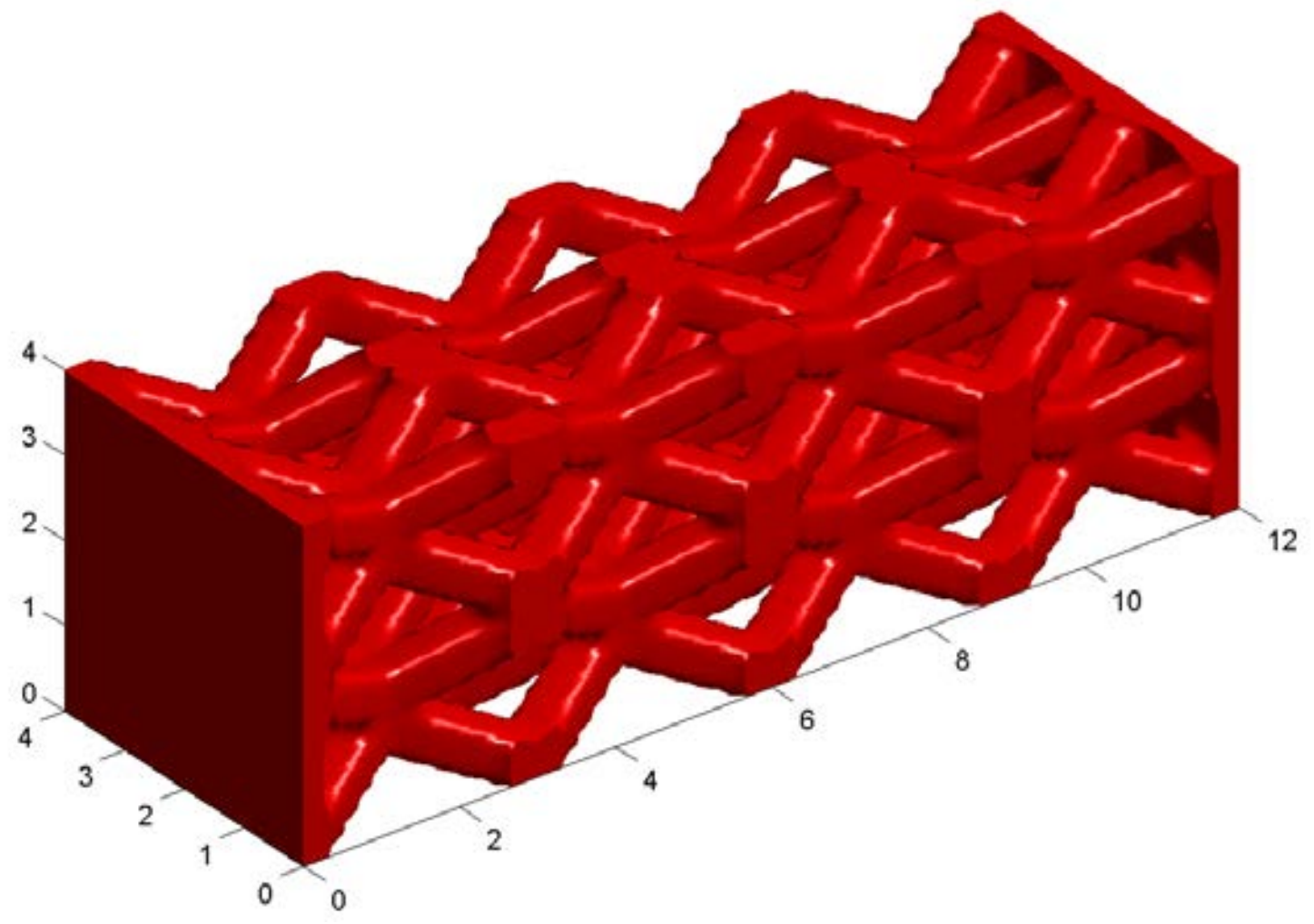

Fig. 23a The initial design for the torsion beam example by the MMC approach.

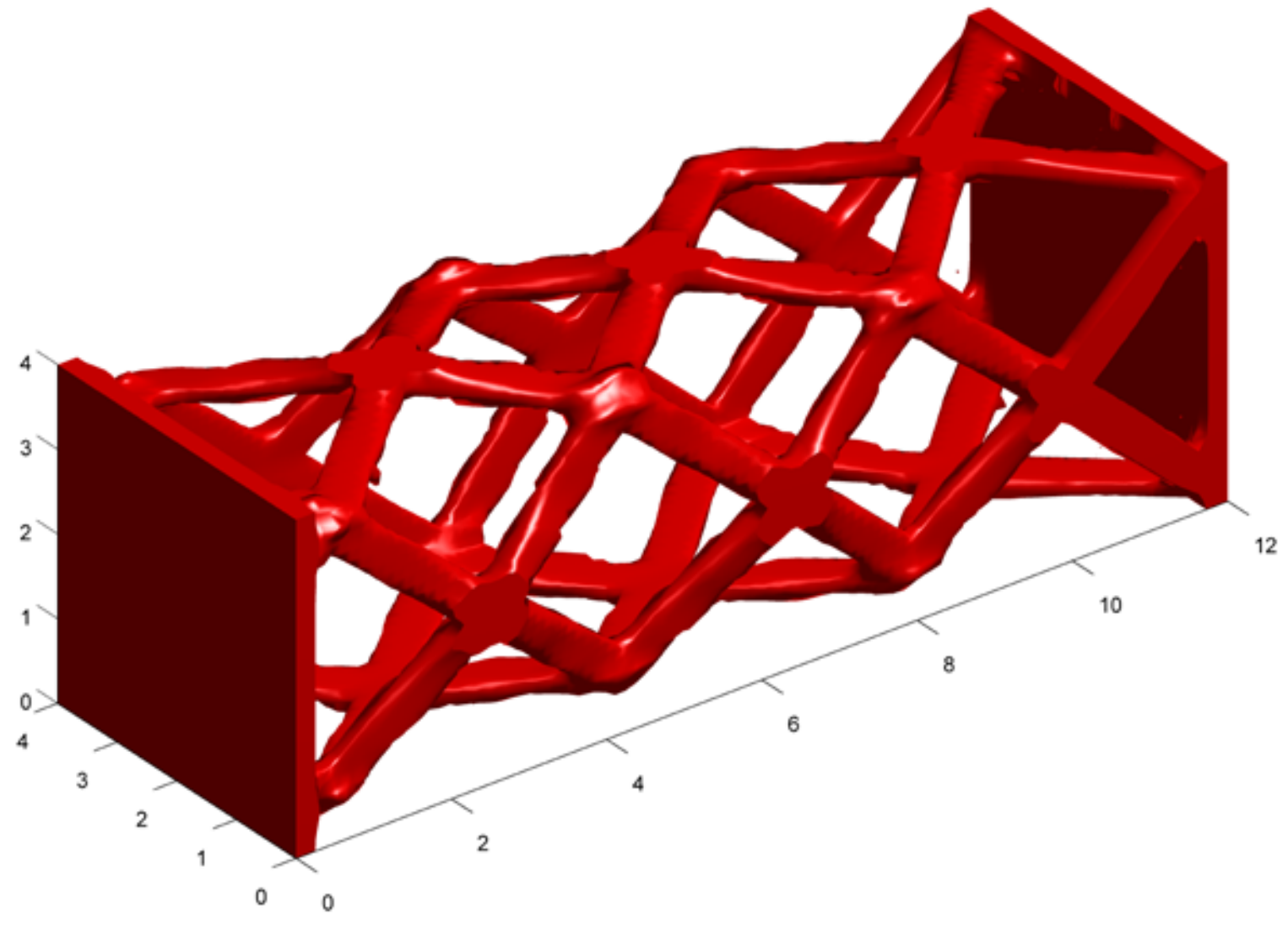

Fig. 23b Optimized structures for the torsion beam example obtained by the MMC approach.

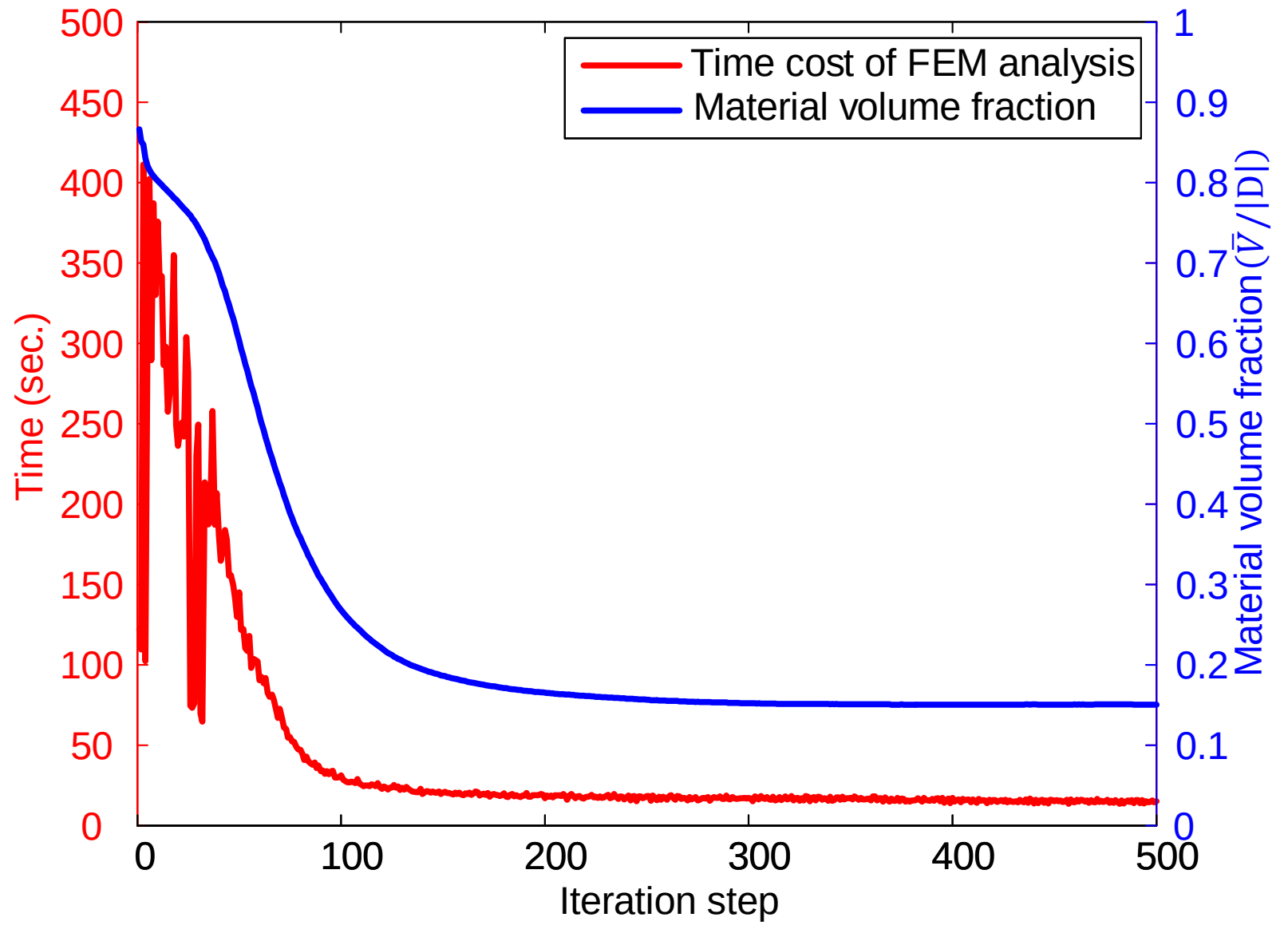

a. The case of the MMV approach.

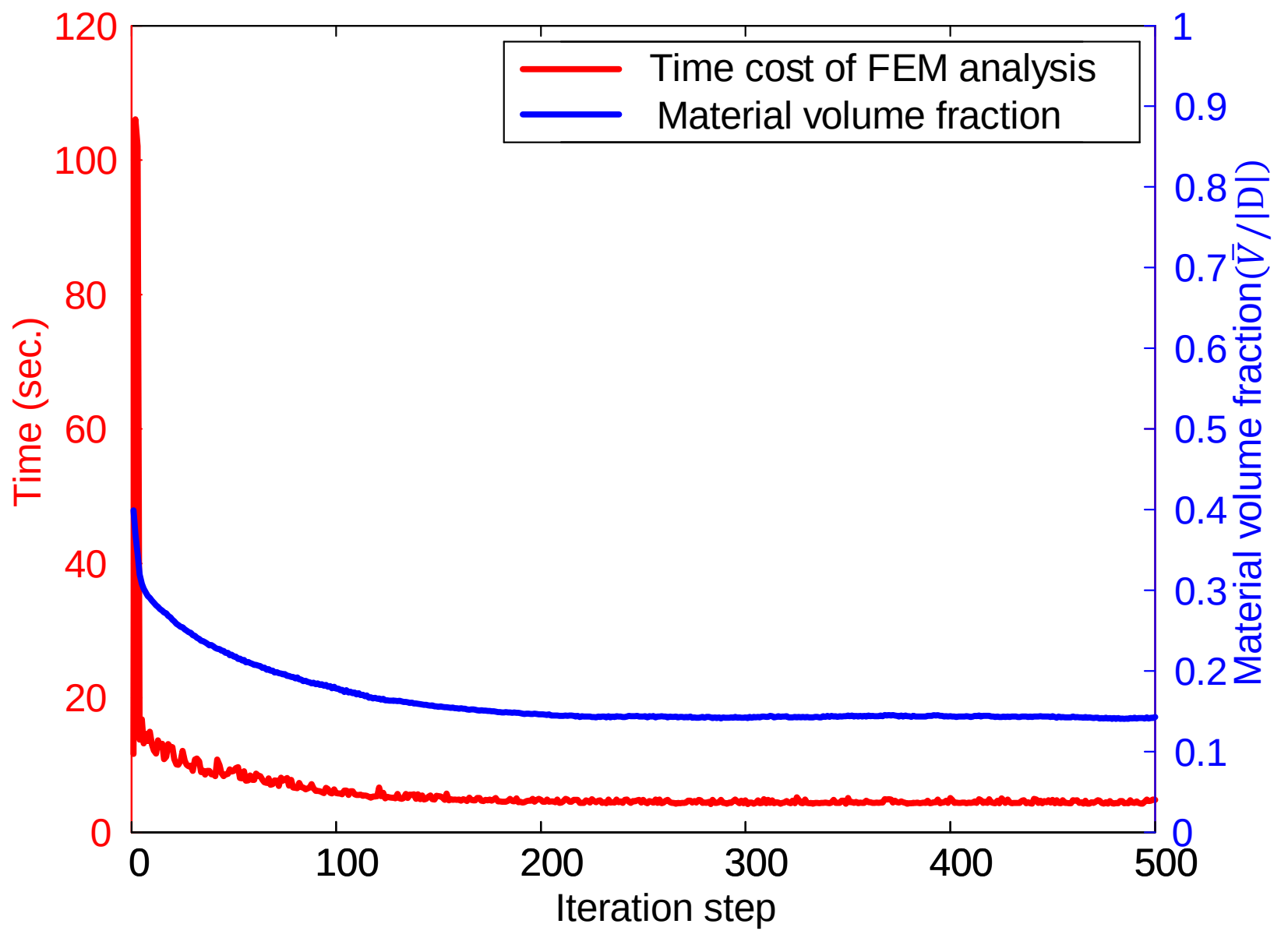

b. The case of the MMC approach.

Fig. 24 The CPU time for the FE analysis of the torsion beam example.

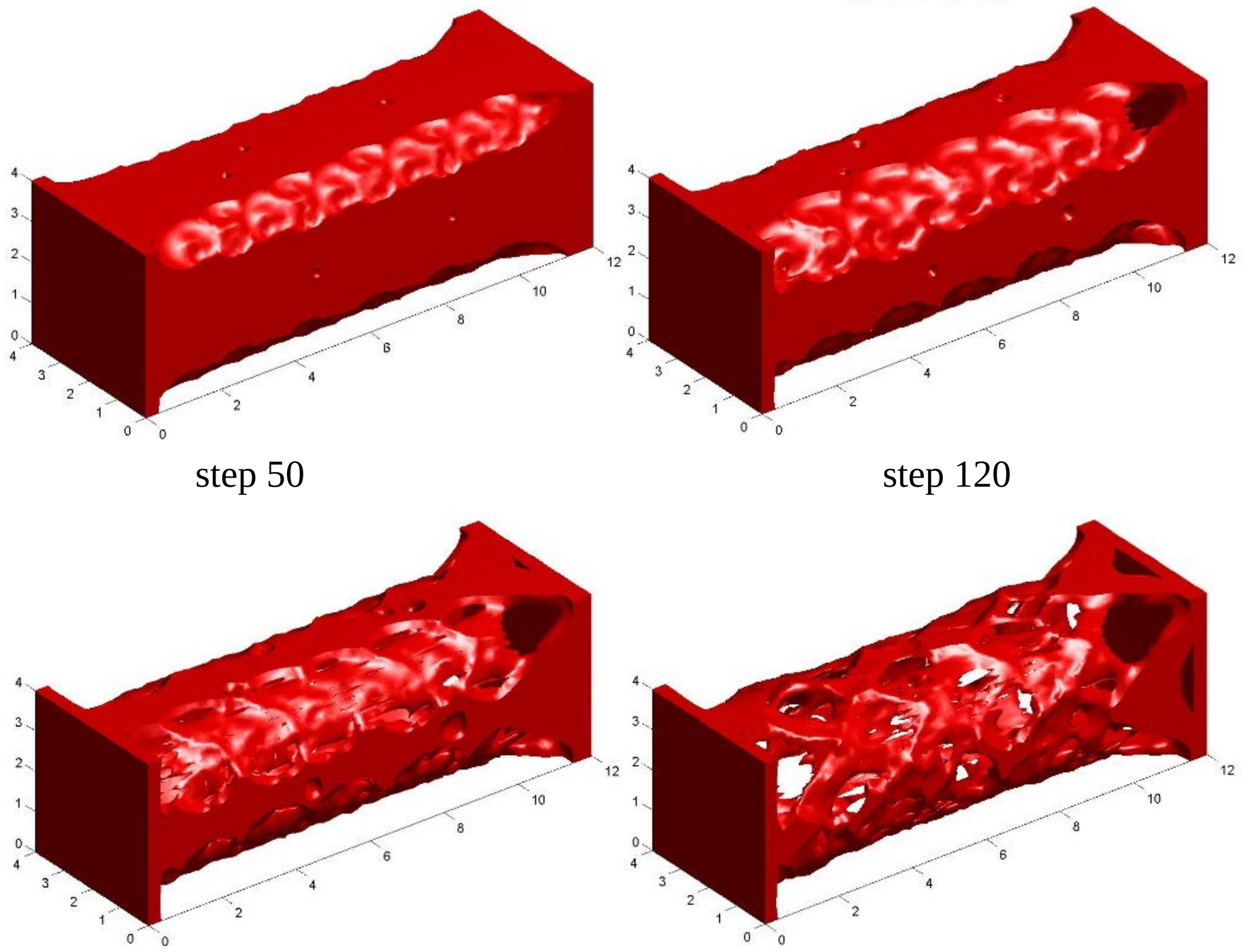

step 50 step 120

step 200 step 400

a. The case of the MMV approach.

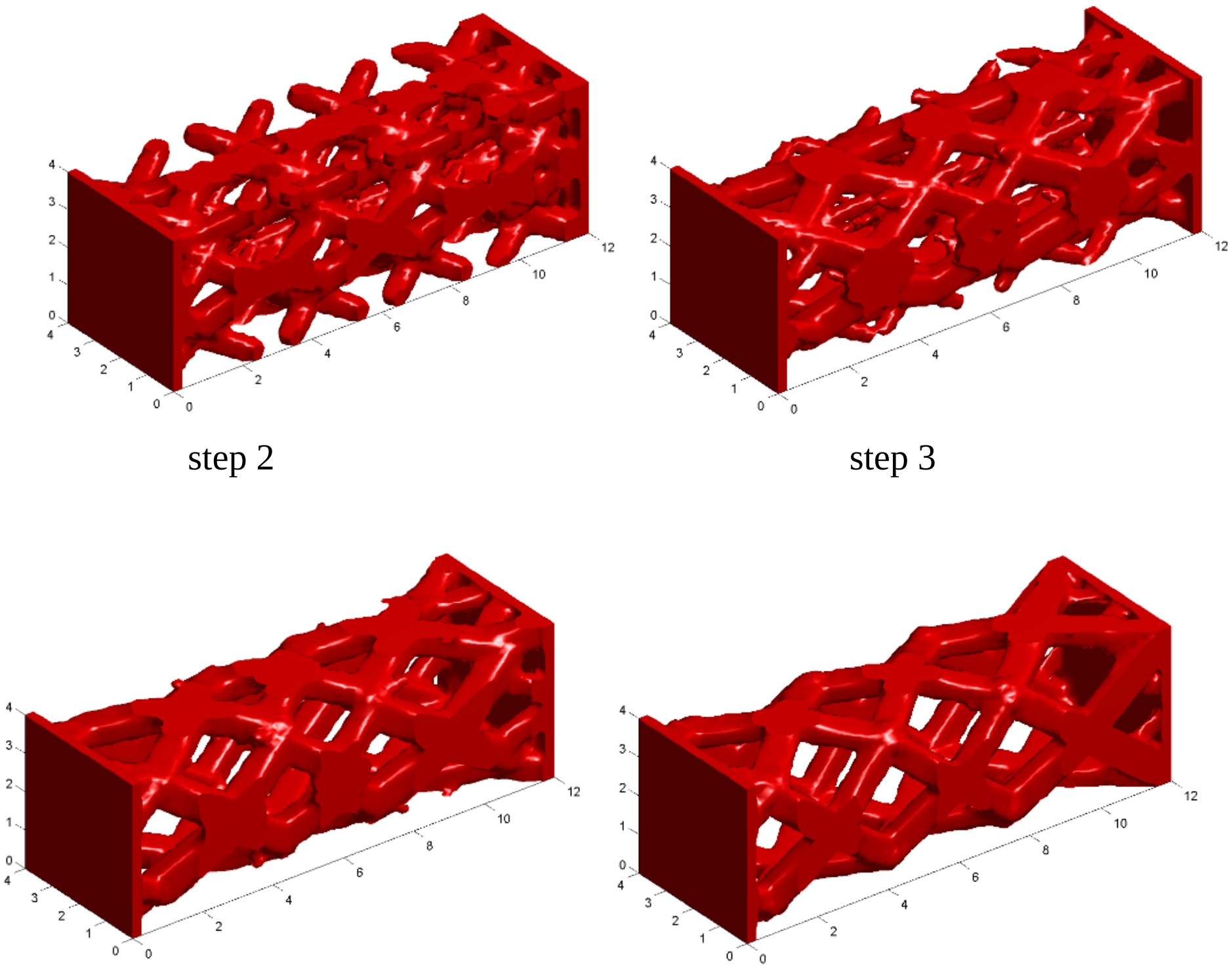

step 2 step 3

step 5 step 20

b. The case of the MMC approach.

Fig. 25 Some intermediate iteration steps for the torsion beam example.